\documentclass[preprint2]{pp7}

\input{pp7.h}

\usepackage{times}
\usepackage{color}

\usepackage{savesym}
\savesymbol{captionbox}
\usepackage{caption}
\restoresymbol{CAPTIONBOX}{captionbox}

\usepackage{mhchem}

\usepackage[switch]{lineno}

\usepackage{hyperref}

\begin{document}

\title{\textbf{\LARGE Chemical Habitability:\\Supply and Retention of Life’s Essential Elements During Planet Formation}}

\author {\textbf{\large Sebastiaan Krijt}}
\affil{\small\em School of Physics and Astronomy, University of Exeter, Stocker Road, Exeter EX4 4QL, UK (\href{mailto:s.krijt@exeter.ac.uk}{s.krijt@exeter.ac.uk})}
\author {\textbf{\large Mihkel Kama}}
\affil{\small\em Department of Physics and Astronomy, University College London, Gower Street, London, WC1E 6BT, UK}
\affil{\small\em Tartu Observatory, University of Tartu, Observatooriumi 1, T\~{o}ravere, 61602, Estonia}
\author {\textbf{\large Melissa McClure}}
\affil{\small\em Leiden Observatory, Leiden University, 2300 RA Leiden, the Netherlands}
\author {\textbf{\large Johanna Teske}}
\affil{\small\em Earth and Planets Laboratory, Carnegie Institution for Science, 5241 Broad Branch Road, NW, Washington, DC 20015, USA}
\author {\textbf{\large Edwin A. Bergin}}
\affil{\small\em Department of Astronomy, University of Michigan, 323 West Hall, 1085 S. University Avenue, Ann Arbor, MI 48109, USA}
\author {\textbf{\large Oliver Shorttle}}
\affil{\small\em Institute of Astronomy \& Department of Earth Sciences, University of Cambridge, UK}
\author {\textbf{\large Kevin J. Walsh}}
\affil{\small\em Department of Space Studies, Southwest Research Institute, 1050 Walnut St., Boulder, CO 80302, USA}
\author {\textbf{\large Sean N. Raymond}}
\affil{\small\em Laboratoire d'Astrophysique de Bordeaux, Univ. Bordeaux, CNRS, B18N, All. Geoffroy Saint-Hilaire, 33615 Pessac, France}

\begin{abstract}
\baselineskip = 11pt
\leftskip = 0.65in 
\rightskip = 0.65in
\parindent=1pc
{\small
Carbon, Hydrogen, Nitrogen, Oxygen, Phosphorus and Sulfur (CHNOPS) play key roles in the origin and proliferation of life on Earth. Given the universality of physics and chemistry, not least the ubiquity of water as a solvent and carbon as a backbone of complex molecules, CHNOPS are likely crucial to most habitable worlds. To help guide and inform the search for potentially habitable and ultimately inhabited environments, we begin by summarizing the CHNOPS budget of various reservoirs on Earth, their role in shaping our biosphere, and their origins in the Solar Nebula. We then synthesize our current understanding of how these elements behave and are distributed in diverse astrophysical settings, tracing their journeys from synthesis in dying stars to molecular clouds, protoplanetary settings, and ultimately temperate rocky planets around main sequence stars. We end by identifying key branching points during this journey, highlighting instances where a forming planets' distribution of CHNOPS can be altered dramatically, and speculating about the consequences for the chemical habitability of these worlds.
 \\~\\~\\~}
\end{abstract}  

\section{\textbf{INTRODUCTION}} \label{sec:intro}
Almost three decades have passed since the discovery of the first exo-planet around a main sequence star. Children born after the publication of \citet{Mayor95} are now finishing their PhDs in rapidly evolving and maturing fields like astrobiology and exoclimatology. It will be up to them to use the next generation of ground-based and space-based observatories to search for and identity existing biospheres on nearby planets and provide a satisfactory answer to the question that has been on everyone's mind since 1995 (and before): \emph{Are we alone in the universe?}

This next generation of scientists has evolved from organisms that first appeared on Earth perhaps within a few 100 Myr of the Moon-forming impact, arguably the last step in the assembly of Earth some 4.56 Gyr ago. All life on Earth requires a solvent (water), energy (light from the Sun and redox chemical energy), and a series of bio-essential elements (carbon, hydrogen,  nitrogen,  oxygen,  phosphorus,  sulfur -- hereafter ``CHNOPS'' -- to construct molecules and polymers with a range of shapes, properties, and uses \citep{Baross2020,Sasselov2020}. Given the abundance of these elements in the Galactic disk, it can be argued that life originating elsewhere will make use of similar ingredients \citep{Hoehler2020}.

To help guide and inform the search for potentially habitable and ultimately inhabited environments, it is then useful to consider where these similar ingredients can come together as the result of planet formation and planet evolution \citep{DesMarais2008,ZahnleCarlson2020}. A lot of attention has been given to the (classical) Habitable Zone \citep{Kasting1993}, the region around a star where a planet (with some assumed properties and atmospheric make-up) could host liquid water on its surface. Even though the boundaries depend on the properties of said hypothetical planet \citep[e.g.][]{Kopparapu2013}, the habitable zone is primarily an attribute of the central star. See \cite{Kaltenegger2017} for a recent discussion of what properties of stars and mature planets define their habitable zones.

Here, we synthesize the state-of-the-art in understanding how universal cosmic processes form worlds with potentially habitable compositions. We use the term \emph{chemical habitability}, an inherent property of planets distinct from the largely externally-constrained classical habitable zone concept. A world that is chemically habitable has: 1) a supply of carbon, hydrogen, nitrogen, oxygen, phosphorus, sulfur ("CHNOPS"), and other bio-essential elements that are accessible to prebiotic chemistry \citep{Rimmer2019, Sasselov2020}, and 2) is capable of maintaining the availability of the CHNOPS elements over relevant geologic timescales. So defined, the chemical habitability of a world is built upon the distribution and cycling of its CHNOPS elements; features of a planet that are conditional upon the earliest phases of planet formation, right through to its present tectonic state. 

State-of-the art remote sensing observations of increasingly smaller exo-planets can be combined to provide information about their bulk densities, atmospheric properties, and orbital parameters. Architectures of multi-planet systems can be studied and compared, and planet occurence rates can be estimated and improved, especially for the most easy-to-spot planets (Currie et al.; Lissauer et al.; Weiss et al., this volume). However, at present challenges arising from the presence of clouds and hazes \citep[e.g.,][]{Crossfield&Kreidberg2017,Gao2021,Barstow2021}, stellar activity \citep[e.g.,][]{dumusque2017, Rackham2018, CollierCameron2018,Cegla2019, Iyer&Line2020, Mayorga2021}, and degeneracies associated from fitting interior models to a single bulk density \citep[e.g.,][]{Dorn2015,Grimm2018,Plotnykov&Valencia2020} prevent the detailed characterisation of the CHNOPS budget on/in terrestrial planet analogs.

Important additional context comes from an increasingly sophisticated view of how planetary systems form. Much of the progress since \emph{Protostars \& Planets VI} has been driven by observations. Apart from the discovery of new exoplanet classes that had to be explained (Hot Jupiters, Mini-Neptunes, etc.), facilities like Herschel and ALMA have revolutionized our understanding of the reservoirs of atoms, molecules, and dust grains in protostellar and protoplanetary systems on spatial scales comparable to the size of the solar system (Manara et al.; Miotello et al.; Benisty et al., this volume). Indirect (Pinte et al., this volume) and direct \citep{Keppler2018,Haffert2019} evidence of accreting giant planets embedded in young gas-rich disks has been found. High-resolulation observations of dozens of systems can be arranged to outline evolutionary sequences \citep{Garufi2018,Cieza2021} and compared to exoplanet demographics \citep{vanderMarelMulders2021,Mulders2021}. Driven by these observations, new ideas in planet formation theory have rapidly been developed and embraced. Modern planet formation theory now presents an unbroken chain of processes connecting the raw materials (e.g. dust, gas) in protostellar systems to fully-formed planets (e.g. Dr{\c{a}}{\.z}kowska et al., Lesur et al., Paardekooper et al. this volume).

In some cases, findings may appear to alienate our Earth and Solar System from the rest of the Galaxy. For example, cold Jupiters are fairly rare, even amongst FGK stars \citep{Raymond2020}, and super Earths, the planet type most common in the Galaxy \citep{Fulton2017}, is absent here. At the same time, many perhaps unexpected connections between astronomical observations/theory and Solar System studies have appeared: the idea of Jupiter separating material reservoirs in the early solar nebula \citep{Kruijer17} resembles closely the planetary interpretation of abundant substructures seen with ALMA \citep{Andrews2020}; convincing evidence for the streaming instability route to planetesimal formation \citep{Johansen2014} was found in our own Kuiper Belt \citep{Nesvorny2019,McKinnon2020}; and pebble accretion models are now frequently invoked to explain the rapid growth of the Solar System's giant \citep{Johansen2017,Alibert2018} and even terrestrial planets \citep{Johansen21}. In-person visits from interloping interstellar planetesimals/comets like 1I/'Oumuamua \citep{Meech2017} and 2I/Borisov \citep{JewittLuu2019} are perhaps the most poetic examples of overlap between planet formation at home and abroad.

We begin this review of chemical habitability by discussing the distribution and origin of CHNOPS on Earth (Sect.~\ref{sec:Earthbackintime}), and the origin and variation of CHNOPS in other stars (Sect.~\ref{sec:CHNOPS_COSMOS}). In Sect.~\ref{sec:exoplanet_hosts} we highlight observed links between exo-planet properties and host star properties related broadly to chemical habitability, and in Sects.~\ref{sec:3clouddisk} and \ref{sec:planet_formation} discuss the varied behavior of CHNOPS en route to and during planet formation. Finally, in Sect.~\ref{sec:branching_points} we summarize and assess what we view as key branching points in the formation of chemically habitable terrestrial planets. Our focus is on temperate rocky worlds near the habitable zones of (single) main sequence stars, but as we will see understanding these planets' CHNOPS budgets requires a holistic treatment of the formation of entire planetary systems.

\section{\textbf{TRACING THE EARTH'S INGREDIENTS BACK THROUGH TIME}}\label{sec:Earthbackintime}

The Earth provides a key constraint on how planets may attain chemical habitability.  Here is a planet where, in principle, we can do the accounting: CHNOPS fluxes between reservoirs can be measured, and the reservoirs themselves probed directly, or indirectly, for budgetary estimates (Sect.~\ref{subsec:earth_chnops}).  Reaching further back in time, fossil evidence places earliest life as being older than $\rm3.5\,Ga$ \citep{schopf2006_rsoc}, with more controversial observations from the rock record and molecular clock analysis placing it even earlier, perhaps 3.7${-}\rm4.1\,Ga$ \citep{bell2015_pnas,nutman2016_nature,betts2018_neco}.  These older ages push the emergence of life into the Hadean eon and by necessity then also the point in time when Earth's chemical habitability was established.  Only a few hundred million years earlier than this was the moon forming impact \citep{jacobson2014_nature,maurice2020_sciadv}, and Earth's history grades into that of the wider solar system (Sect.~ \ref{subsec:accreting_earth}), as the remnants of planet formation were swept up by the larger planets.  This is a key period in which the earlier CHNOPS losses, from the tenuous atmospheres of precursory planetary embryos, may have been made up for in the `late accretion' of undifferentiated (i.e., only mildly heated) objects.   Earlier still and Earth's history fragments into its innumerable building blocks and ultimately back to the nebular processes that set the stage for planet formation (Sect.~ \ref{subsec:protosolarnebula}).

\begin{deluxetable}{l l l l l l l l l l l l}

\centering

\tabletypesize{\scriptsize}
\tablewidth{0pt}
\tablecaption{The C-H-N budget of the Earth$^1$}
\tablehead{ \colhead{ Reservoir$^2$ } &  \multicolumn{2}{c}{ Reservoir mass} & \multicolumn{3}{c}{C} & \multicolumn{3}{c}{H}  & \multicolumn{3}{c}{N}\\
\colhead{} &  (kg) &  (M$_\oplus$) & (kg) & \%C$_\text{BE}$ & (M$_\mathrm{atm}$) & (kg) & \%H$_\text{BE}$ & (M$_\mathrm{ocean}$) & (kg) & \%N$_\text{BE}$ & (M$_\mathrm{atm}$)}

\startdata
Atmosphere     & $5.2\times10^{18}$ & $10^{-6}$ & $8.8\times10^{14}$ & $10^{-5}$ & $1$      & $1.3\times10^{14}$ & $10^{-5}$  & $10^{-6}$ & $4.0\times10^{18}$ & $14$  & $1$ \\
Biosphere      & $1.1\times10^{16}$ & $10^{-9}$ & $5.5\times10^{15}$ & $10^{-4}$ & $6.3$    & $2.8\times10^{12}$ & $10^{-7}$ & $10^{-8}$ & $9.7\times10^{14}$ & $10^{-4}$ & $10^{-4}$ \\
Hydrosphere    & $1.6\times10^{21}$ & $10^{-4}$ & $3.8\times10^{17}$ & $10^{-2}$ & $430$    & $1.8\times10^{20}$ & $14$      & 1         & $2.4\times10^{16}$ & $10^{-2}$ & $10^{-2}$\\
Crust          & $2.0\times10^{22}$ & $10^{-3}$ & $7.1\times10^{19}$ & $1$       & $10^5$   & $3.5\times10^{19}$ & $3$      & $0.19$    & $2.3\times10^{18}$ & $1$ & $0.58$\\
Lithosphere    & $1.4\times10^{23}$ & $0.02$    & ${\approx}6.0\times10^{19}$ & $1$ & $10^{5}$ & ${\approx}3.9\times10^{17}$ & $10^{-2}$ & $10^{-3}$ & ${\approx}6.0\times10^{18}$ & $2$ & $1.5$\\
Shallow mantle & $9.4\times10^{23}$ & $0.16$    & $3.5\times10^{19}$ & $10^{-1}$ & $10^{4}$ & $1.6\times10^{19}$ & $1$       & $0.09$    & $7.2\times10^{18}$ & $3$ & $1.8$\\
Deep mantle    & $2.9\times10^{24}$ & $0.49$    & ${\approx}2.9\times10^{21\dagger}$ & $39$ & $10^{6}$ & ${\approx}3.3\times10^{20\dagger}$ & $26$      & $1.8$     & ${\approx}1.7\times10^{19\dagger}$ & $6$ & $4.3$ \\
Core           & $1.9\times10^{24}$ & $0.33$    & ${\approx}4.4\times10^{21\dagger}$ & $59$ & $10^{7}$ & ${\approx}7.2\times10^{20\dagger}$ & $56$     & $4$       & ${\approx}2.4\times10^{20\dagger}$ & $87$ & $60$\\
\multicolumn{12}{l}{} \\

\multicolumn{12}{l}{\emph{Totals} } \\
BSE & $4.03\times10^{24}$ & 0.675 & $3.1\times10^{21}$ & 41 & $3\times10^{6}$ & $5.6\times10^{20}$ & 44 & $3$ & $3.7\times10^{19}$ & 13 & $9$\\
Bulk Earth          & $5.97\times10^{24}$ & 1     & $7.5\times10^{21}$ & & $9\times10^{6}$ & $1.3\times10^{21}$ & & $7$ & $2.8\times10^{20}$ & & $70$\\
\enddata

\tablenotetext{1}{\scriptsize{BSE = Bulk silicate Earth. ${\approx}$ indicate mass estimates with significant uncertainty, $^{\dagger}$ indicating those that also impact our understanding of the total CHNOPS budget of the Earth, and by implication the processes that will distribute the elements among planetary reservoirs more generally. References for compiling reservoir estimates, C: \citet{bar2018_pnas,sleep2001_jgr,sleep2009_g3,levoyer2017_ncomms,Marty12,dasgupta2013_revmin}; H: \citet{lecuyer1998_chemgeol,hirschmann2012_phys,wu2018_jgr}; and N: \citet{johnson2015_esr}}\label{tab:chnops_earth1}}

\tablenotetext{2}{\scriptsize{The crust represents the sum of the oceanic and continental crust.  The lithosphere is the sub-continental lithospheric mantle, i.e., the long-lived and non-convecting part of the mantle reservoir.  The `shallow mantle' represents the convecting mantle sampled by mid-ocean ridge volcanism, the mantle above the mantle transition zone at $\rm670\,km$ depth.  The `Deep mantle' is then all the silicate Earth below the transition zone.  The bulk silicate Earth is all terrestrial reservoirs outside of the core, i.e., that are ultimately accessible to life and the surface environment.}}

\normalsize
\end{deluxetable}

\begin{deluxetable}{l l l l l l l l l l}
\tabletypesize{\scriptsize}
\centering
\tablewidth{0pt}
\tablecaption{The O-P-S budget of the Earth$^1$}
\tablehead{ \colhead{ Reservoir } & \multicolumn{3}{c}{O$^2$} & \multicolumn{3}{c}{P} & \multicolumn{3}{c}{S}\\

\colhead{} & (kg) & \%O$_\text{BE}$ & (M$_\mathrm{atm}$) & (kg) & \%P$_\text{BE}$ & (M$_\mathrm{ocean}$) & (kg) & \%S$_\text{BE}$ & (M$_\mathrm{atm}$)}

\startdata
Atmosphere     & $1.2\times10^{18}$  & $10^{-4}$  & $1$       & $-$                 & $-$    & $-$      & $4.8\times10^{9}$  & $10^{-11}$ & $10^{-9}$ \\
Biosphere      & $4.4\times10^{15}$  & $10^{-6}$  & $10^{-3}$ & $1.3\times10^{14}$  & $10^{-6}$ & $1$      & $1.1\times10^{13}$  & $10^{-8}$ & $10^{-5}$ \\
Hydrosphere    & $2.5\times10^{18}$  & $10^{-3}$  & $2$       & $1.3\times10^{14}$  & $10^{-6}$ & $1$      & $1.3\times10^{18}$  & $10^{-3}$ & $1$ \\
Crust          & $8.2\times10^{19}$  & $10^{-2}$  & $70$      & $6.3\times10^{18}$  & $10^{-1}$ & $10^{4}$ & $8.4\times10^{18}$  & $10^{-2}$ & $6.5$ \\
Lithosphere    & ${\approx}4.8\times10^{19}$  & $10^{-2}$  & $40$      & ${\approx}9.1\times10^{18}$  & $10^{-1}$ & $10^{5}$ & $2.1\times10^{19}$  & $10^{-1}$ & $16$ \\
Shallow mantle & $3.9\times10^{20}$  & $10^{-1}$  & $300$     & $4.6\times10^{19}$  & $1$ & $10^{5}$ & $1.4\times10^{20}$  & $10^{-1}$ & $100$\\
Deep mantle    & ${\approx}1.2\times10^{21}$  & $10^{-1}$  & $10^3$    & $3.0\times10^{20}$  & $6$ & $10^{6}$ & $8.4\times10^{20}$  & $3$ & $600$ \\
Core           & $-4.8\times10^{23}$ & $10^{2}$ & $-10^{5}$ & ${\approx}4.9\times10^{21\dagger}$  & $93$ & $10^{7}$ & ${\approx}2.9\times10^{22\dagger}$  & $97$ & $10^{4}$ \\
\multicolumn{10}{l}{} \\

\multicolumn{10}{l}{\emph{Totals} } \\
BSE & $1.7\times10^{21}$  & $10^{-1}$ & $1\times10^{3}$  & $3.6\times10^{20}$ & 7 & $3\times10^{6}$ & $1\times10^{21}$ & 3 & $800$ \\
Bulk Earth          & $-4.8\times10^{23}$ &  & $-4\times10^{5}$ & $5.2\times10^{21}$ & & $4\times10^{7}$ & $3\times10^{22}$ & & $2\times10^{4}$ \\
\enddata

\tablenotetext{1}{\scriptsize{BSE = Bulk silicate Earth. ${\approx}$ indicate mass estimates with significant uncertainty, $^{\dagger}$ indicating those that also impact our understanding of the total CHNOPS budget of the Earth, and by implication the processes that will distribute the elements among planetary reservoirs more generally. References for compiling reservoir estimates, O: \citet{lecuyer1999_epsl,workman2005_epsl,hirose2013_annrev,McDonough95}; P: \citet{rudnick2014_treatise,white2014_treatise,rudnick1998_chemgeol,workman2005_epsl,McDonough95}; and S: \citet{brimblecombe2013_treatise,rudnick2014_treatise,lorand2003_gca,McDonough95,hirose2013_annrev}}\label{tab:chnops_earth2}}

\tablenotetext{2}{\scriptsize{Oxygen is counted here as oxygen required to move the system to the reference redox state (e.g., following \citet{evans2006_geology}); when oxygen \emph{removal} is required this is counted as a positive number (as the system contains oxygen compared to the reference state) and oxygen addition being counted as a negative number (as the system is lacking oxygen compared to the reference state).  For the atmosphere the reference state is defined as an oxygen-free atmosphere; for the biosphere the reference state is oxygen-free organic matter; for the hydrosphere this is dioxygen-free seawater (i.e., counting the small amount of dissolved oxygen in seawater) and neutral sulfur (counting the large amount of oxygen in sulfate); for the silicate portions of the planet (crust, lithosphere, convecting mantle, and core), we focus on the oxidation state of Fe, as the dominant multi-valent element.  The reference state we choose is \ce{Fe^{2+}O}, with reservoirs either being more oxidsed if they contain \ce{Fe^{3+}}, or reduced if they contain Fe-metal (\ce{Fe^0}).  The core's negative oxygen abundance occurs because the reducing power of its metallic iron overwhelms the small amount of oxygen it may contain.}}

\normalsize
\end{deluxetable}

\begin{figure*}[t]
    \includegraphics[clip=,width=.9\textwidth]{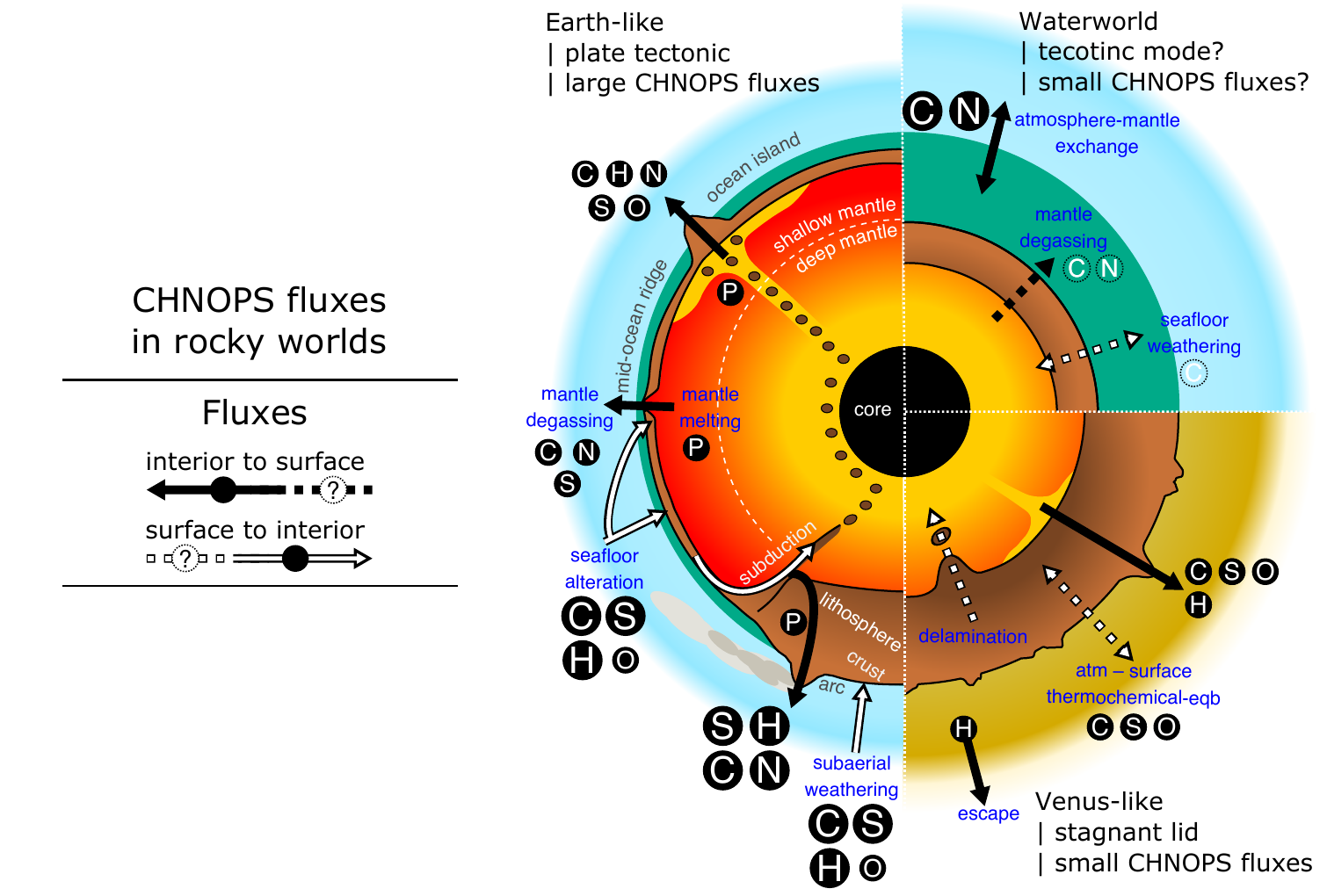}
\caption{Abiotic CHNOPS fluxes on mature rocky worlds.  Left, fluxes on an Earth-like planet, with plate tectonics, a hydrosphere, and a low-mass atmosphere. Top right, fluxes on a water world planet.  Bottom right, fluxes on Venus-like planets, with high surface temperatures, dry atmospheres, and stagnant lid tectonics.  Black arrows indicate fluxes from the interior to surface, white arrows indicate fluxes from the surface to the interior, black symbols with element names indicate the elements involved in the process, with larger circles indicative of larger fluxes.  Dotted arrows and hollow symbols indicate uncertain fluxes (e.g., for many of the CHNOPS fluxes on water worlds).\label{f:chnops_fluxes}}
\end{figure*}

\subsection{The importance and distribution of CHNOPS on Earth\label{subsec:earth_chnops}}

Much of what makes Earth the precious `pale blue dot' \citep{sagan1997_pbd} can be attributed to the CHNOPS elements.  Clearly these elements are of central importance to the biosphere -- at an elemental level these elements \emph{are} the biosphere.  However, CHNOPS also play a fundamental role in the abiotic dynamics of the modern Earth, from processes at the top of the atmosphere into our planet's core.  This is evident from Tables \ref{tab:chnops_earth1} \& \ref{tab:chnops_earth2}, which list estimates for the budgets of CHNOPS in terrestrial reservoirs from the surface atmosphere-biosphere-hydrosphere, into the planet's silicate crust-lithosphere-mantle, and finally in the core.  With the exception of P in the atmosphere, CHNOPS are pervasively distributed throughout the Earth system, and are capable of having a major impact on planetary dynamics at even part per million concentrations (e.g., as with carbon and greenhouse warming of the atmosphere).  The chemical properties of CHNOS (less so P) distinguish them from the major Fe-O-Mg-Si-Ca-Al elemental building blocks of the Earth (where O is a special case; \citealt{McDonough95}), which are bound in mineral form.  Properties of the biotic and abiotic Earth therefore have a common dependence on CHNOPS, a dependence that is likely to be replicated on habitable and inhabited planets throughout the Galaxy \citep[e.g.,][]{cockell2021_astrobiology}.  In this section we review some of the non-biological planetary processes CHNOPS are involved in that are key for setting planetary habitability (section \ref{sss:why_chnops}), before moving onto the distribution and history of CHOPS redistribution on Earth (section \ref{sss:where_chnops}).  

\subsubsection{Why being `Earth-like' requires CHNOPS\label{sss:why_chnops}}

Earth is unique in a solar system context and, given what we know presently, in an exoplanetary context (Lissauer et al., this volume).  Indeed, our knowledge of Earth is so detailed and specific that the concept of `Earth-like' requires generalisation to be of any use to exoplanetary sciences \citep{shorttle2021_elements}. The minimalist definition captured by $\eta_\oplus$ in \citet{bryson2020_mnras} (orbital period and radius within 20\% of Earth's) implies maybe 1 in 50 GK stars host an `Earth'.  However, a planet being 'Earth-like' still allows much room for non-Earth-like conditions on and inside the planet, differences that will very likely be rooted in a planet's divergent history of CHNOPS gain and distribution.  Here, we emphasise two standout aspects of the modern Earth, its geodynamic mode of plate tectonics and its carbon cycle, phenomena habitable planets may be universally required to possess \citep{ehlmann2016_jgr}, and which both are intimately coupled to CHNOPS.

\emph{Plate tectonics:} This is an elegant theory that emerged in the mid-twentieth century to describe the long-term motions of Earth's surface, that the surface moves as rigid `plates', being destroyed at trenches (subduction zones) and formed at mid-ocean ridges \citep{mckenzie1967_nature,morgan1968_jgr}.  Despite this apparent simplicity, plate tectonics is the beating heart of our planet, connecting the evolution of the deepest mantle to the atmosphere and biosphere.  It does this by enabling the cycling of huge masses of material, $\sim 10^{15}\,\rm{kg\,yr^{-1}}$, through mid-ocean ridge melting regions and subduction zones, allowing CHNOPS fluxes out of (at mid-ocean ridges and arc volcanoes) and into (at subduction zones) the Earth's interior (Fig. \ref{f:chnops_fluxes}).  This plate tectonic flux of material is large enough that, if sustained over $4.5\,\rm{Gyr}$, it will process the entire silicate portion of the planet.  Crucially for habitability, plate tectonics allows CHNOPS elements to be removed from the surface environment for hundred-million-year timescales by sequestration into our planet's interior, \emph{and} allows for their restoration by degassing during mountain building and magmatism (Fig. \ref{f:chnops_fluxes}).  This provides a mechanism whereby accumulation of e.g., \ce{CO2}, \ce{O2}, \ce{SO2}, and \ce{H2O} at the Earth's surface can be regulated by their drawdown into mineral form for longterm storage, but with the potential for re-release if their reservoir is subsequently embroiled in mountain building or magmatism: high temperature processes that drive volatile elements like C, H, N, O, and S, and elements like P with chemical affinity for melts, back to the surface.

The cycling of CHNOPS, sustained by plate tectonics on Earth, may be very different under other planetary geodynamic regimes (Fig. \ref{f:chnops_fluxes}, left vs. right).  On stagnant lid planets, of which modern Venus is likely an example, the thick lithosphere (the non-convecting `lid' of the planet) can only exchange with the interior by blobbing off at the bottom (`delamination'), whilst volcanic degassing from the interior is suppressed. Limited volcanic degassing on Venus is evidenced by the planet's atmospheric argon (Ar) inventory. Argon can trace the time integrated degassing of planetary mantles because \ce{^{40}Ar} is produced \emph{in rocks} by \ce{^{40}K} decay: as K is a lithophile element present in a planet's rocks, the \ce{^{40}Ar} decay product is also trapped there unless the rocks melt and that melt is transported to the surface.  Therefore, the degree to which the \ce{^{40}Ar} in a planet has been able to escape to the atmosphere constrains the efficiency of melting of the planet's interior.  Venus's atmosphere is observed to contain less radiogenic \ce{^{40}Ar} than Earth's atmosphere, which given an inferred K abundance in the planet suggests it has experienced less volcanic input to its atmosphere than Earth \citep{kaula1999_icarus}.  

Combined, lower rates of volcanism and inefficient recyling of surface material to the planet's interior may limit the efficiency of long-term chemical cycling on stagnant lid planets.  Although modelling has suggested stagnant lid planets may be able to sustain habitability for Gyr timescales \citep[e.g.,][]{tosi2017_aa,foley2018_astrobio}, the effectiveness of the processes involved in this remain highly uncertain.  Further study of Venus, and new constraints on its early history, offer a prime opportunity for solar system sciences to inform exoplanet habitability \citep{Kane2021}.  

Whilst the above suggests that plate tectonics is important for chemical habitability (i.e., CHNOPS), Venus also exemplifies how CHNOPS may be important for plate tectonics.  A central aspect of plate tectonic theory is that the deformation associated with the creation and destruction of the plates is localised.  The epitome of localised deformation is the fault, those fractures in the Earth's crust responsible for seismicity that allow two blocks of crust to slide past each other.  In the absence of water, faults would have to move at their dry frictional strength of over $\rm700\,MPa$ \citep{amiguet2012_epsl}.  This stress is so large that plate tectonics could be entirely suppressed, resulting in a single unbroken plate (a stagnant lid).  However, the presence of water is able to decrease fault strength, both forming low-friction hydrous mineral phases and hydraulically opening the fault \citep{amiguet2012_epsl,sleep1992_nature}.  Hence, plate tectonics has a role in sustaining surface water inventories, and is itself sustained by the presence of water.  This connection between water and tectonics reaches all the way to Earth's core and the creation of the magnetic field, which on Earth-mass planets is likely enabled by plate tectonics's effective extraction of heat from the planet's interior \citep{nimmo2002_geology} --- with the magnetic field helping preserve a planet's surface liquid water inventory by offering enhanced protection against solar wind-driven water loss \citep[e.g.,][]{lundin2007_ssr}.   

\emph{Carbon cycle:} Earth is the only rocky planet in the solar system with a functioning carbon cycle (sketched in Fig. \ref{f:chnops_fluxes}). On Venus, carbon released from the planet's interior is on a one way journey to its atmosphere, where it has so far accumulated ${\sim}90\,\rm{bar}$ of \ce{CO2}, whilst on Mars, the tenuous atmosphere is so cold \ce{CO2} condenses and is unable to support temperate conditions.  Although Earth's carbon cycle is heavily influenced by biology, and may have been so for billions of years, carbon is likely central to achieving environmental homeostasis even on abiotic worlds \citep{Kasting1993}.  A key insight exoplanets can provide is how effective this abiotic carbon cycle is \citep{lehmer2020_ncomm}, whether it is indeed effective in the absence of biology, and how robust it is to changing planetary regimes, e.g., the water-worlds, which may be common outcomes of planet formation (Fig.~\ref{f:chnops_fluxes}; \citealt{kite2018_apj,foley2018_astrobio,Lichtenberg2019}).  

However, the carbon cycle is not just reliant on a planet's inventory of C.  The geochemical cycle removing carbon from the atmosphere requires abundant liquid water to enable the silicate weathering reaction, in which 2 moles of \ce{CO2} dissolve to produce carbonic acid, which reacts with rocks, and ultimately sequesters one mole of that \ce{CO2} in mineral form as carbonate \citep{urey1952_pnas}.  The sensitivity of this reaction to temperature, via kinetics and rainfall \citep{walker1981_jgr,maher2014_science}, creates homeostatic feedback whereby increasing temperatures driven by \ce{CO2} greenhouse forcing increases the rate of reactions that remove \ce{CO2} from the atmosphere.  Though \ce{CO2}-\ce{H2O} are at the centre of climate regulation on the modern Earth, it is also important to note the direct climate role of other CHON species (but likely not P or S). Early in its history Earth was faced with a `faint young sun' problem, in that the ${\sim}70\%$ weaker light from the young sun would have been insufficient, given Earth's present atmosphere, to maintain liquid water at its surface \cite{sagan1972_science}.  The abundance of CHON gas species in Earth's early atmosphere is central to solving this problem and reconciling the geological record of liquid water with climate models.  High \ce{CH4} and \ce{CO2} concentrations and/or a combination of \ce{N2}-\ce{H2} warming may have sustained habitable conditions on the early (possibly pre-biotic) Earth \citep{catling2020_sciadv,wordsworth2013_science}.

\subsubsection{Earth's CHNOPS through space and time\label{sss:where_chnops}}

To understand what is required for chemical habitability it is essential to know the distribution of CHNOPS on the modern Earth.  The budgets of major planetary reservoirs are listed in Tables \ref{tab:chnops_earth1} \& \ref{tab:chnops_earth2}, and below we discuss the key factors governing their distribution and redistribution.

\textit{Carbon: }Most of Earth's carbon is likely in its core, which results from carbon acting as a siderophile element at the high pressures at which metal-silicate segregation took place on Earth \citep{dasgupta2013_revmin}: i.e., carbon has a strong affinity for metallic phases rather than silicate oxide phases in the extreme conditions of planetary interiors.  However, this carbon is inaccessible during subsequent planetary evolution, so the core's key role occurs early by setting the proportion of the total planetary \ce{C} inventory available for subsequent distribution and cycling among terrestrial reservoirs.  In `accessible', i.e., silicate Earth, the overwhelming majority of carbon is stored in the mantle as carbonate (shallow mantle), metal-sulfur alloys \citep{zhang2019_epsl}, and diamond (deep mantle; \citealt{dasgupta2013_revmin}). The amount present in the deep mantle has large uncertainties \citep{miller2019_epsl,Marty20}.  Mantle carbon is supplied to the atmosphere when melting occurs; carbon has a strong affinity for silicate liquids over minerals \citep{rosenthal2015_epsl}, meaning it enters magmas and the magma migrates to the surface where carbon's volatile nature drives it into the gas phase and thereby the ocean-atmosphere system.  The efficacy of silicate weathering in removing this volcanic carbon from Earth's atmosphere is evidenced by the very small amount of Earth's carbon inventory that the atmosphere contains and the large crustal carbon pool it overlies: $10^5$ times the present atmospheric inventory, an amount that if put back into the atmosphere would give Earth a Venus-like $\rm{\sim}100\,bar$ \ce{CO2} atmosphere (Table \ref{tab:chnops_earth1}).

\textit{Hydrogen: }Unlike for carbon, the surface environment is a major reservoir of hydrogen on Earth, where it is present in oxide form as water.  Earth has been able to retain surface water over its history, unlike Venus, because of the stratospheric cold trap that prevents water molecules reaching a height in the atmosphere where photodissociation can produce H atoms that are light enough to escape to space.  Water interacts with the crust and produces hydrous mineral phases with structurally bound hydrogen, which have formed a significant reservoir of Earth's surface water (${\sim}15\%$) being locked in the crust \citep{lecuyer1998_chemgeol}.  It has been proposed that such reactions could provide a major (and terminal) sink for water on some planets \citep{wade2017_nature}, but on Earth the restoration of water occurs by degassing at volcanoes and during metamorphism (high temperature/pressure transformation of rocks; Fig. \ref{f:chnops_fluxes}).  Earth's lithospheric and mantle reservoirs combined contain between 1 and 2 Earth oceans of hydrogen \citep{hirschmann2012_phys}, primarily stored as defects in the crystal lattice.  A very significant fraction of Earth's hydrogen could also be present in the core \citep{wu2018_jgr,tagawa2021_ncomms}, although significant uncertainty remains in estimating its core abundance. 

As with carbon, hydrogen has an affinity for silicate liquids over silicate minerals \citep[e.g.,][]{peslier2017_ssr}, meaning mantle melting transports water from the planet's interior to the surface.  Unlike carbon, however, there are strong feedbacks between water storage in the mantle and the vigour of mantle convection, as hydrogen defects in minerals have a profound impact on the viscosity of `dry' rocks. Numerical simulations have shown how in principle these feedbacks can regulate the surface inventory of water on Earth, with increased mantle water lowering mantle viscosity, increasing the vigour of convection, which drives more mantle melting and thereby drives water back out of the mantle \citep{mcgovern1989_epsl}.  These strong feedbacks between water storage in the mantle and the processes that cycle it between Earth's interior and surface may explain the observation that Earth's oceans have maintained an approximately constant volume through time \cite[e.g.,][]{windley1977_nature}.  

\begin{figure*}[t]
\centering
    \includegraphics[clip=,width=0.9\textwidth]{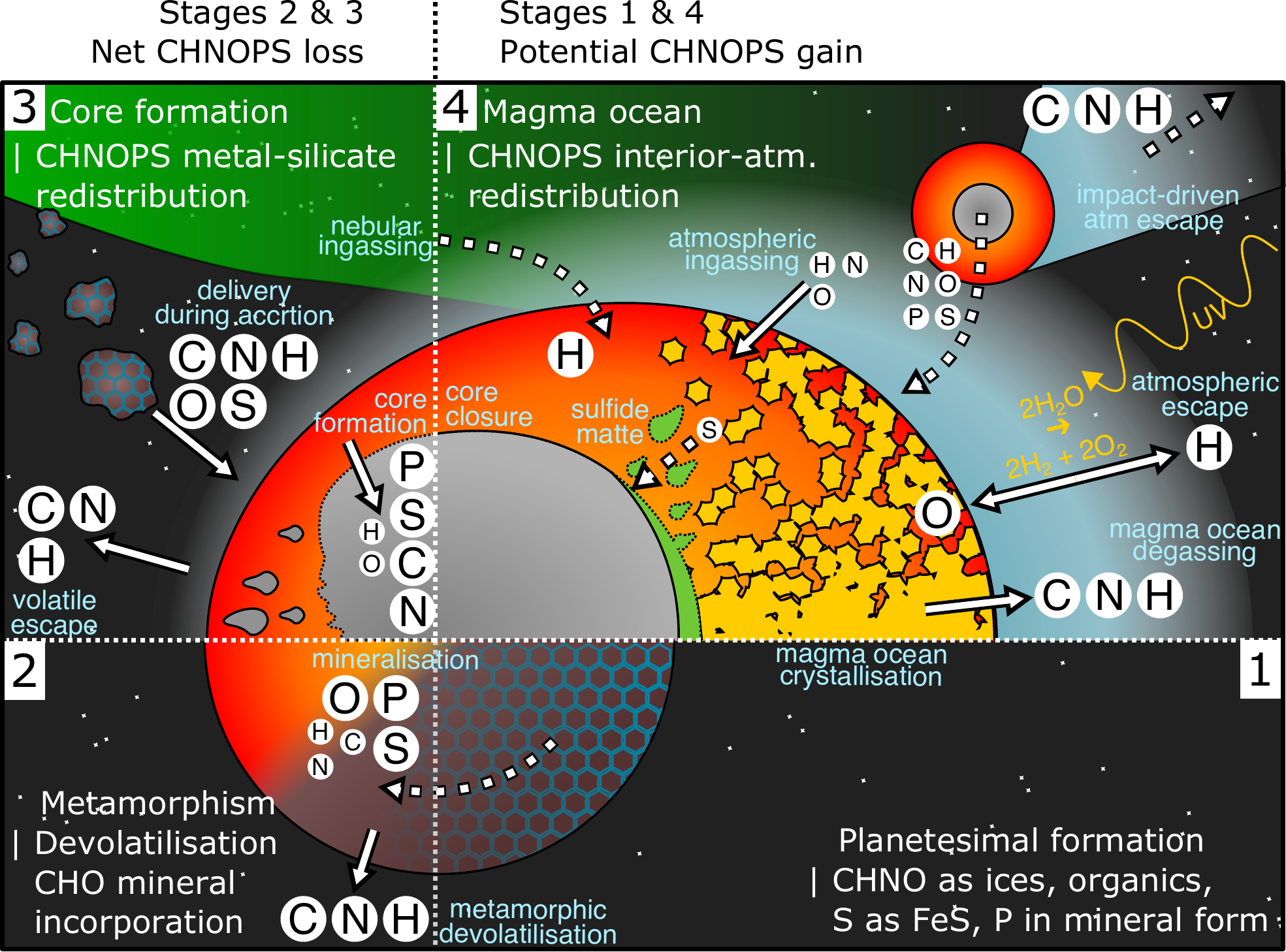}
\caption{CHNOPS fluxes during planetary assembly.  CHNOPS are tracked through the planetesimal assembly stage (1), planetesimal heating by short lived radionuclide decay driving metamorphic volatile processing and loss (2), eventual planetesimal melting and growth (3), and magma ocean crystallisation on Earth-sized objects (4).  Stages (2) and (3) represent periods of potentially significant CHNOPS loss, either to space or the object's growing metallic core.  Whilst in stage (4) the planet is massive enough to retain C-N-O-P-S against atmospheric escape (H is likely still lost at this stage), further loss of siderophile elements to the core is prevented as it becomes isolated from the silicate portion of the planet, C-N-H are being expelled from the crystallising magma ocean into the nascent atmosphere, and S may be substantially lost to the lower mantle and outer core during sulfide matte segregation.\label{f:evo_dist}}
\end{figure*}

\textit{Nitrogen: }Has major reservoirs in the atmosphere, crust, mantle and core \citep{johnson2015_esr}.  The presence of so much nitrogen in the planet's interior is at first surprising, because nitrogen's chemical similarity to the noble gases might suggest it should have primarily entered Earth's atmosphere during the magma ocean epoch and have had little route back into the interior thereafter.  However, nitrogen's solubility in silicate melts has been shown to be strongly dependent on oxygen fugacity (a measure of how oxidising a system is), allowing for a reducing early atmosphere to have driven nitrogen into Earth's magma ocean (Fig. \ref{f:evo_dist}; \citealt{libourel2003_gca,wordsworth2016_epsl}).  Crucially, this model predicts that the distribution of nitrogen on a planet is tied closely to how oxidised the magma ocean atmosphere is, which will depend on the abundance and rate of oxidising power created in the atmosphere by water dissociation and hydrogen loss to space \citep{wordsworth2016_epsl} --- a prediction that may eventually be testable from exoplanet observations.    An important additional sink for N during the planet formation stage is the core, which may host a majority of Earth's N incorporated into interstices in the Fe-metal alloy (\cite{li2016_gpl}; Table \ref{tab:chnops_earth1}).  As with atmosphere-magma ocean partitioning, experimental work has shown that metal-silicate nitrogen partitioning is sensitive to how oxidising the system is at the time of equilibration, with more oxidised conditions promoting nitrogen storage in the core \citep{grewal2019_gca}.  There are therefore two important events in a planet's life, during core formation and during magma ocean-atmosphere equilibration, where the fate of its nitrogen may be decided by how oxidising its surface and/or interior is.

As nitrogen is the major constituent of Earth's atmosphere, archives of paleo-atmospheric pressure indirectly probe its accumulation in the surface environment.  Various attempts have been made to infer past atmospheric pressure using fossilised raindrops \citep{som2012_nature}, the size of bubbles frozen in magmas \citep{som2016_ngeo}, the oxidation state of micrometeorites \citep{rimmer2019_gpl}, and the noble gas composition of trapped bubbles of gas \citep[e.g.,][]{avice2018_gca}.  Combined, these proxies suggest paleo-atmospheric pressures in the Earth's Archean ($\rm{>}2.5\,Ga$) less than or equal to the present day.  If anything then, nitrogen may have been added to Earth's atmosphere over time, which has significant implications for mechanisms of early climate warming, which cannot then rely on \ce{N2}-based collisionally-induced absorption \citep{wordsworth2013_science}.

\textit{Oxygen: }Is unique amongst the CHNOPS elements in not being a minor element in the bulk Earth, rather it is the most abundant element in the planet \citep[e.g.,][]{palme2014_treatise}.  However, \emph{free} oxygen (as dioxygen) is vanishingly rare in a planetary context, with most oxygen bound as metal oxides, and even slight variations in its chemical availability driving significant changes in, for example, atmospheric chemistry, the composition of volcanic gasses, and the physical properties of the Earth's mantle. In Table \ref{tab:chnops_earth2}, oxygen accounting is done by crudely considering `available' oxygen, that which is present in excess of a reference state (apart from in the atmosphere, where just \ce{O2} is considered).  In most of the silicate Earth we simply consider \ce{Fe^{2+}O} as this reference oxide, and therefore account for available oxygen as that present in \ce{Fe^{3+}O_{1.5}}.  In the core, conversely, iron metal (\ce{Fe^0}) provides a potential oxygen sink (for both atmospheric and mantle oxygen), and contributes `negative' oxygen to our budgeting.  Apparent from Table \ref{tab:chnops_earth2} is that the core's reducing power overwhelms all the available oxygen in all terrestrial reservoirs combined --- the bulk Earth is very reduced \citep[e.g.,][]{zhang2017_lpsc}.  However, Earth's shallower reservoirs are saved from being tied to the reducing conditions of the core, as at least since the time of magma ocean crystallisation the core has been effectively chemically isolated from the silicate Earth.  Whilst Table~\ref{tab:chnops_earth2} shows that there is in principle many times the atmospheric inventory of `available' oxygen in the mantle, the mantle itself is much more reducing than Earth's modern atmosphere, which with 21\% \ce{O2} is extremely oxidising.  Therefore, chemical interaction between the mantle and atmosphere would lead to oxygen transfer from the atmosphere into rocks; this is indeed the process that happens on Earth to form the long term oxygen cycle \citep{lecuyer1999_epsl,shorttle2015_epsl}.  In this context, the fragility of Earth's 21\% atmospheric oxygen is evident, as it sits atop a vast and vastly more reducing reservoir of rock and iron.  This unstable state is only sustained by oxygenic photosynthesis, with dioxygen's rapid loss from the atmosphere inevitable were the biosphere to collapse \citep{lecuyer1999_epsl,ozaki2021_ngeo}. 

It is important to note that for chemical habitability free oxygen is likely to be highly problematic: all life on Earth exploits reduced forms of carbon that are destroyed in the presence of oxygen, and the chemistry to start life would be even more vulnerable to oxidative destruction \citep{Sasselov2020}.  Therefore, by agnostic (less anthropocentric) standards of habitability, the modern Earth is a profoundly hostile environment.  Fortunately, Earth's surface has not always been like this, and before the Great Oxidation Event at $\rm\sim2.3\,Ga$ ago the atmosphere was almost completely free of \ce{O2} \citep{holland2002_gca,catling2020_sciadv}.  There has been significant debate over whether the timing of the great oxidation event was due to an evolution in the oxidising properties of mantle-derived magmas and gases  \citep[e.g.,][]{holland2002_gca}, versus having been triggered by the biological innovation of oxygenic photosynthesis \citep[e.g.,][]{fischer2016_annrev}.  From a planetary perspective, it is interesting to note that while Earth's mantle rocks are reduced compared to its atmosphere, it has the most oxidised mantle in the solar system \citep[with there being a lack of constraint on Venus's mantle; ][]{wadhwa2008_revmin}, and has a low iron content that diminishes its absolute buffering capacity.  Whilst these facts must have contributed to allowing biological \ce{O2} production to overwhelm reducing sinks of oxygen $\rm{\sim}2.3\,Ga$ \citep{lecuyer1999_epsl,sleep2005_met}, there is little evidence presently for long-term changes in how reducing Earth's mantle has been \citep[e.g.,][]{trail2011_nature}.  

\textit{Phosphorus: }Likely constitutes the limiting nutrient for life on Earth \citep[e.g.,][]{tyrrell1999_nature}.  The throttling of the biosphere by phosphorus availability results from its absence as a stable gas phase in the atmosphere (unlike CHNO), meaning life must access its required P in aqueous form.  Aqueous environments obtain their supply of P from rock weathering, however, the low solubility of phosphate (\ce{PO3^{2-}}) in water means P released from rocks rapidly re-enters mineral form \citep{ruttenberg2003_treatise}.  Life therefore goes to remarkable efforts to recycle the pool of P it has available to it, with the result that in some environments almost all P is contained in biomass \citep{ruttenberg2003_treatise}.  Huge reservoirs of P exist in the silicate Earth, either as a P-bearing phase such as apatite (\ce{Ca10(PO4)6(OH,Cl,F)2}, as in the crust), or as a trace component in nominally P-free mineral phases (as is the case in the mantle).  However, in all these reservoirs, whether the crust, mantle, or mantle-derived magmas, the P concentration in rocks is typically low, at $\rm{\sim}0.1\,\text{wt}\%$.  As observed for C and N, P is also moderately siderophile \citep[e.g.,][]{righter2018_gca}, meaning significant loss occurred to Earth's core (Table \ref{tab:chnops_earth2}). 

The major mineral reservoir that life relies on for its P supply may have changed significantly over Earth's history.  On the modern Earth most crustal P is bound in the P-mineral apatite.  However, the presence of apatite in rocks depends on their composition, and the less silica-rich nature of early crustal rocks may have limited its presence \citep[e.g.,][]{lipp2021_gpl}.  Instead, the early P cycle may have had to rely on trace P liberated from basalts \citep{walton2021_esr}.  Stepping back further in time, meteoritic P sources may have been essential in supplying reactive P for prebiotic chemistry, in the form of schreibersite \citep[\ce{(Fe,Ni)3P}, ][]{walton2021_esr}.  

\textit{Sulfur: }Is degassed from the mantle at volcanoes, primarily as \ce{SO2} and \ce{H2S} \citep[e.g.,][]{symonds1993_ajs}.  However, these species are unstable in Earth's atmosphere, lasting just days--weeks before they are deposited out or converted to sulfuric acid.  For this reason the modern oceans host huge reserves of sulfur as sulfate, ultimately sourced from oxidation of the reduced volcanic sulfur species \citep{brimblecombe2013_treatise}.  This oceanic sulfur reservoir cycles back to the mantle via subduction of sulfide minerals, produced in part by biological sulfate reduction, and sulfate minerals, precipitated during hydrothermal circulation at mid-ocean ridges and in evaporite deposits.  In the mantle, sulfur is stored as sulfide minerals/melts with distinct physical properties compared to the silicate minerals.  This difference potentially came into play most dramatically during Earth's magma ocean solidification when saturation of the crystallising magma in sulfide drove segregation of a separate sulfide liquid (the `Hadean matte'), which due to its high density sank to the core \citep[Fig. \ref{f:evo_dist}; ][]{oneill1991_gca}.  This episode has enormous implications not only for the remaining budget of S on Earth, but also of other sulfide-loving (`chalcophile') elements \citep[e.g., silver; ][]{righter2019_maps}.  Sulfide matte formation is sensitive to the chemical conditions that affect sulfide solubility in magma, its pressure, temperature, and composition.  Notably, the loss of chalcophile elements to a sulfide matte is not thought to have occurred on Mars \citep{righter2019_maps}, and sulfur on the planet is inferred to be correspondingly more abundant and more important in surface geochemical cycles than on Earth \citep[e.g.,][]{king2010_elements}.  As well as being lost at a late stage by sulfide matte formation, sulfur is siderophile, partitioning into the core during metal segregation (\ref{f:evo_dist}).  Together, these processes explain the prediction that Earth's core hosts a majority of the planet's sulfur (Table \ref{tab:chnops_earth2}).

The major transition in sulfur behaviour at Earth's surface occurred in response to the Great Oxidation Event.  Before this time, sulfur was stable at Earth's surface as sulfide minerals, and sulfate concentrations in the oceans were extremely low.  After the rise of atmospheric oxygen at $\rm{\sim}2.3\,Ga$, sulfide minerals were no longer stable, being oxidised to sulfate on exposure to the atmosphere and hence ushering in significant sulfate concentrations in Earth's oceans \citep[e.g.,][]{brimblecombe2013_treatise}. 

\subsection{The Accretion of Earth}\label{subsec:accreting_earth}

Bulk Earth's chemical and isotopic composition was determined by both {\em how} it grew and the materials from which it grew. The Earth has important contributions of materials from potentially widely separated reservoirs. We review briefly key constraints pertaining to the the timing of accretion, composition and dynamical origin of the Earth's building blocks.

\subsubsection{Timing Earth's Accretion}
When discussing timing, we equate $t=0$ to the time of CAI formation (calcium aluminum inclusions, the oldest nebular condensates with an age of 4.567 Gyr, \citealt{Connelly08,Connelly2012}). The gaseous solar nebula dispersed about 4-5 Myr after the formation of CAIs \citep{Wang2017}. By this time, Mars was pretty much fully formed \citep{DauphusPourmand2011}, and both Venus and Earth were large enough (${\gtrsim}0.5M_\oplus$) to bind a primordial atmosphere H/He \citep{Marty12, Williams2019}, but small enough (${\lesssim}0.8M_\oplus$) to lose it after the nebula dispersed \citep{WuMohanty2016,Lammer2018}. Another timing constraint is provided by the Moon-forming impact, arguably the most dramatic event in the history of the Earth (and the Moon), which occurred some 50{-}150 Myr after the formation of the solar system, long after the solar nebula had dispersed \citep[e.g.,][]{Kleine09,jacobson2014_nature,Avice2014,ZahnleCarlson2020}.

\subsubsection{Up to the Moon-forming impact}
The Earth is considered to be largely built from material that was isotopically similar (but not identical, see e.g., \citealt{FitoussiBourdon2012}) to Enstatite chondrite (EC) meteorites, which are typically linked to asteroids found on the innermost edge of the asteroid belt \citep{Dauphas2017}. Chemically, however, there are several differences. Earth's upper mantle does not match Enstatite's Mg/Si, and the bulk silicate Earth's (BSE's) C/N, C/H, and C/S ratios differ from ECs as well \citep{DasguptaGrewal2019}. Finally, Enstatite chondrites are generally considered overwhelmingly dry, with some being even drier than Earth \citep{ZahnleCarlson2020}, although \citet{Piani2020} recently inferred water contents of between $0.08{-}0.5\mathrm{~wt\%}$ for a collection of EC meteorites suggesting their parent bodies may have been much wetter than previously thought. Contributions of C and N are generally more similar to carbonaceous chondrite (CC) meteorites that predominate further out in the asteroid belt \citep{Marty12}. In total, comets are responsible for at most 1\% of Earth's water, and at most a few percent the nitrogen, but may have delivered significant amounts of noble gases and organic material \citep{Marty2016,Marty2017,Altwegg19}. The origins of Earth's water are discussed in more detail in \citet{Meech19}.

Dynamically, the accretion from these two reservoirs (EC-like and CC-like) is often tied to the formation and behavior of Jupiter. First, isotope properties of iron meteorites \citep{Budde2016,Kruijer17} and an isotopic bimodality, particularly in titanium and chromium, amongst a wide class of meteorites \citep{Warren2011} point strongly to different reservoirs of carbonaceous type materials and non-carbonaceous type materials with limited inter-mixing during much of the gas disk lifetime (see also chapter by Nomura et al.). This barrier is frequently attributed to the rapid formation of Jupiter \citep{Morbidelli2016,Kruijer17}, although non-planetary origins for this dichotomoty have also been proposed \citep[e.g.][]{Brasser2020}. Recent models of planetesimal formation have found that the isotopic dichotomy may potentially correspond to distinct episodes of planetesimal formation that are separated in space or time \citep[e.g.][]{Lichtenberg2021a,Morbidelli2022}.
Separated from the rest of the nebula, Earth, Mars, and Venus and grew from materials with similar isotopic fingerprints \citep{Burkhardt2021}, while chemical differences arose from disk and planetary processes (see Sect.~\ref{sec:planet_formation}).

The rapid accretion of gas by Jupiter perturbed nearby planetesimal orbits strongly and scattered many inward \citep{Raymond2017}. The gas disk can damp their orbits and provide a capture mechanism, thus providing a generic pathway for outer solar system planetesimals to find their way to the Main Asteroid Belt and also to accrete on to the still growing terrestrial planets. Indeed, the diverse structure of the Main Asteroid Belt \citep{demeo2014} demands significant input of material from the outer Solar System to the inner Solar System in the form of plantesimal-sized objects (30-300 km).

Jupiter's orbital migration may have had a similar effect on the distribution of planetesimals of different compositions, scattering a large enough mass in carbonaceous type planetesimals to account for the mass observed in the asteroid belt and total masses accretion of water on Earth \citep{Walsh2011}. While distinguishing these models from each other and any future models for explaining the composition of the asteroid belt is a worthy exercise, fundamentally the asteroid belt serves as a strong indicator of planetesimal transport that likely occurred during the gas disk phase of the Solar System \cite[see review by][]{RaymondNesvorny2020}.

\subsubsection{Moon-forming impact and late veneer}

Following the dissipation of the gas disk, various dynamical mechanisms are still capable of mixing material, with the Earth-Moon impact \citep{Canup2001} and the diverse structure of the Main Asteroid belt as indications of this process \citep{demeo2014,RaymondNesvorny2020}. In fact, the asteroid belt is rich not only with low-albedo C-type asteroids typically linked with carbonaceous chondrite meteorites, but it also has a smattering of D- and P-type interlopers closely linked with the Jupiter Trojans and the Kuiper Belt \citep{Levison2009,DemeoDtype}. These interlopers are successfully modeled as being captured during the giant planet instability after the solar nebula dispersed \citep{Levison2009}. Capture probabilities are quite low \citep{Nesvorny2013} suggesting that the total transport of primitive and ice-rich Kuiper Belt-like material was orders of magnitude lower than what would have been expected during the transport and implantation of the C-type asteroids into the Main Belt.

Nearly every model for terrestrial planet formation in our Solar System ends with the giant impact phase of growth where tens of Moon to Mars sized planetary embryos violently come together to form a system of planets \citep[see][for reviews]{Morbidelli2012,Raymond2020}. Outside of possible growth entirely by pebble accretion (see \citealt{Schiller2018,Johansen21}), there must be a final step for piece-wise accretion of planetary building blocks. The late-stage giant impacts present an opportunity for material mixing across larger distances than often afforded from localized accretion by planetesimals, as long as planetary embryos and planetesimals formed in a broad disk.

The isotopic similarities between the Earth and Moon \citep[e.g.,][]{Wiechert2001,Zhang2012,Dauphas2014}, have spurred numerous works investigating the likelihood that Theia (the Earth-Moon impactor) originated from near enough the Earth to account for their similar compositions. However, large suites of terrestrial planet formation find that this particularly stochastic phase of growth leads to overlapping feeding zones for Venus, Earth and Mars -- which has a slightly different oxygen isotopic composition \citep{Franchi1999} -- and relatively low probabilities that Theia was formed close enough to the Earth to account for the Earth-Moon isotopic similarities (see \citealt{Kaib2015,Mastrobuono-Battisti2015}).

The abundance of highly-siderophile elements in Earth's mantle -- elements that should presumably have sunk to the core during any energetic impact -- indicate that Earth collided with a small population of leftover planetesimals after its final giant impact, generally considered to be the Moon-forming impact \citep{Day2007,Walker2009}. The amount of material accreted in this late veneer makes up ${\approx}0.5\%$ of the Earth's mass \citep{MorbyWood2015,jacobson2014_nature,Dauphas2017}, and the isotopic composition of (based on molybdenum and ruthenium) implies inner solar system (e.g. Enstatite chondrite-like) material dominated also this component \citep{Fischer-Godde2017}. For various assumptions about the tail end of planetary accretion, the final sweep up and accretion of material by the Earth can be used to date the Earth-Moon impact, where only about ${\approx}0.5\%$ of an Earth mass is typically accreted after $\sim$95~Myr, which would then be the nominal time for the Earth-Moon impact \citep[albeit with large error bars;][]{jacobson2014_nature}. \citet{Sakuraba2021} show how, for a relatively oxidized magma ocean, numerous (small) chondritic impactors can explain the BSE's C/N/H depletions.

Based on experimental results finding that C may become less siderophile in some scenarios while N is unaffected, \citet{Grewal2019moon} propose instead that the C/N ratio in bulk silicate Earth may be explained with a single Mars-sized impactor, and rely on far less contribution from carbonaceous chondrite materials. In this scenario, a singular giant impact -- presumably the Earth-Moon forming impact -- is responsible for the delivery of carbon, nitrogen, and sulfur to the BSE. Compared to the late veneer model this alternative pathway would then require a near-absence of available planetesimals, which would otherwise still be accreted.

\subsection{Origins in the Solar Nebula} \label{subsec:protosolarnebula}

The asteroids, comets, and other planetary building blocks have their origins in the solar nebula, the protoplanetary disk of our sun. Here we summarize key observations/inferences regarding the CHNOPS budgets contained within these building blocks. Motivated by the previous section, we discuss for each element the dominant reservoirs in the inner solar system (as traced by Enstatite chondrites), the outer asteroid belt (Carbonaceous Chondrites), and the outer solar nebula (drawing in particular upon new insights from studies of dwarf planet Ceres, the comet 67P/Churyumov-Gerasimenko, Pluto, and the Kuiper Belt Object Arrokoth). The discussion of CHNOPS in the outer expanses of the Solar system is particularly important for the connection to protostellar and protoplanetary systems (Sect.~\ref{sec:3clouddisk}).

For some time, an important focus has been placed on the origin and delivery of water to terrestrial worlds and the formation of the Earth inside the water iceline \citep{vandishoeck14,Meech19}. However, over the past decade there has been substantial work on the origin and supply of other important volatile elements including carbon/nitrogen \citep{Alexander12, Marty2013, Bergin15, Alexander17, Marty20, Johansen21} and sulfur \citep{Kama19}. These works highlight very important trends within the solar system that go beyond the traditional picture of a water iceline and hint at substantial gradients in the volatile element inventory in primitive materials supplied to terrestrial worlds - i.e. the inner solar system solids were ultimately carbon-, nitrogen- and water-poor.

\subsubsection{Two nebular paradigms: reset \emph{vs.} inheritance}

The thermal \emph{reset} model is commonly used when discussing the inner solar system and meteoritic constraints. In this model, all solids are initially sublimated to their elemental form due to viscous heating in the rapidly accreting disk. As the nebula cools, the volatile content of solids is determined by an equilibrium condensation sequence of successive mineral formation based on the half-mass condensation temperature at a given pressure. In this context, deviations from solar abundances observed in e.g., primitive CI chondrites and Earth itself can be linked to the temperature at which they (or their building blocks) condensed. For example, for a gas of solar composition at a pressure of $10^{-4}\mathrm{~bar}$, the rock-forming elements Mg, Fe, and Si condensation temperatures are around 1350 K (condensing as e.g. forsterite and enstatite, taking roughly 20\% of the oxygen with them), P \& S condense around 1230 and 650 K (as \ce{Fe3P} and \ce{FeS}), while H, C, N, and the leftover O freeze out as ices (e.g., \ce{H2O}, \ce{CH4}, \ce{NH3}) at temperatures below a few 100 K \citep{Lodders2003}. Indeed, elements with low condensation temperatures show the largest deviations when comparing solar nebula abundances with primitive CI chondrites, or even Earth itself \citep[e.g.,][]{Grossman72, Lewis72, McDonough95, Fegley20}.

The \emph{inheritance} model is motivated by the presence of pristine (never heated) ices in outer solar system bodies, such as comets with abundances that correlate with ices observed in the interstellar medium tracing stages prior to planetary birth \citep[][discussed in more detail in Sect.~\ref{sec:3clouddisk}]{Bocklee-Morvan00,Altwegg19, Drozdovskaya19}. Further, chondritic meteorites contain pre-solar grains, highlighting even inner solar system materials escaped a complete reset \citep[e.g.,][]{NittlerCiesla2016}.

The interplay of radial in- and outward transport of solids between the reset and pristine reservoirs is an integral process in our picture of the early solar system, and it is in this context that we discuss each CHNOPS element individually, grouping H \& O together to highlight water.

\subsubsection{Evolving CHNOPS reservoirs}

\begin{figure}[t]
    \includegraphics[clip=,width=.45\textwidth]{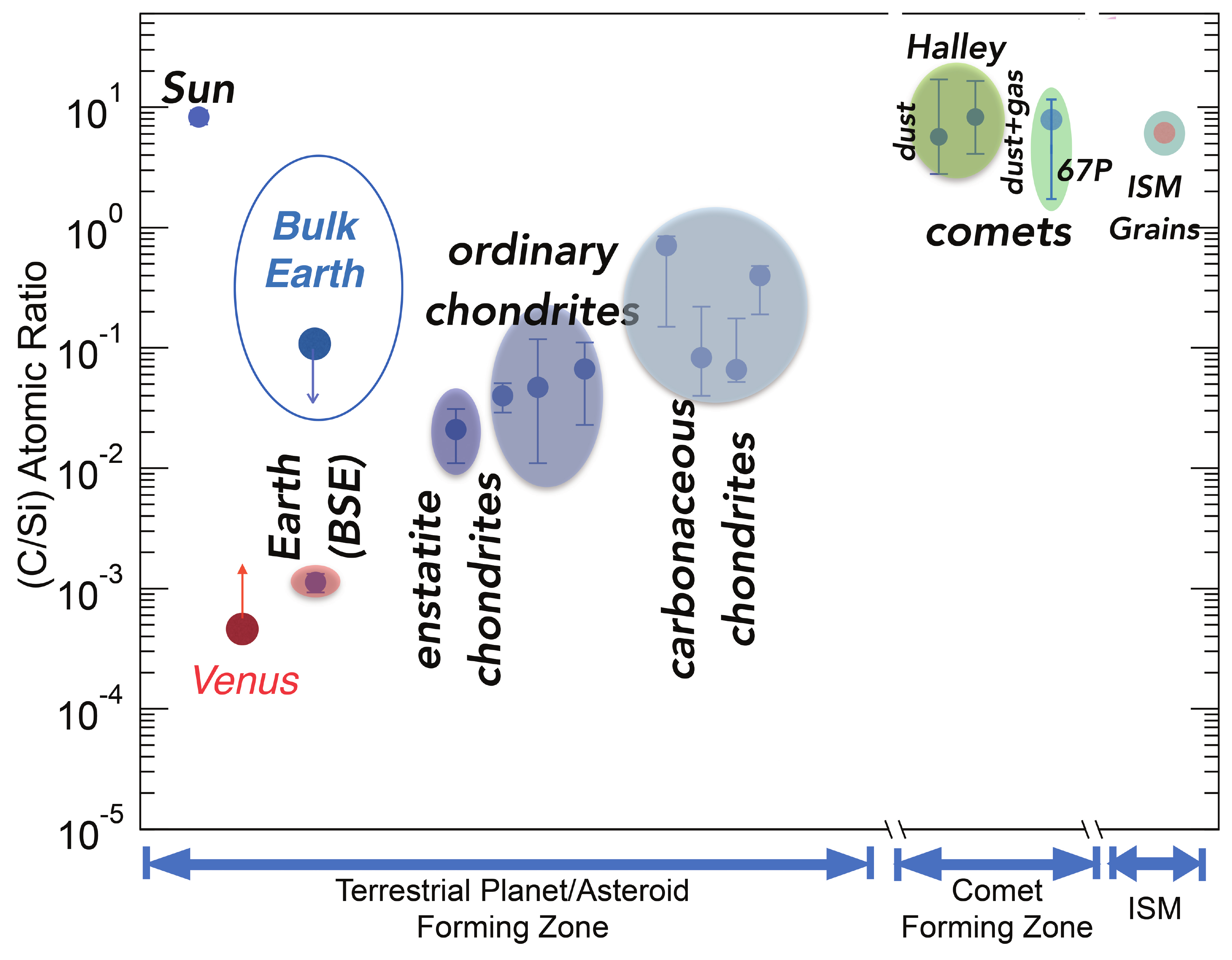}
    \centering
    \caption{Atomic ratios of carbon to silicon in various solar system bodies including the Sun, Earth, Venus, chondrites, comets, and interstellar grains.  Two values are shown for the Earth: the Bulk Silicate Earth (BSE - mantle and fluid envelopes minus the core) and the Bulk Earth (including the stringent upper limit on the amount of carbon in the core).  For chrondrites the error range reflects the intrinsic range within the meteoritic classes, while the errors for cometary composition are estimated uncertainties including measurement error.  For Venus the estimate is from \citet{Halliday13} and is lower limit as the abundance of carbon in the rockly planetary interior is unknown.  The carbon content of 67P is taken from \citet{Rubin19}.
    Figure adapted from \citet{Bergin15} and \citet{Li21}.}
    \label{f:cvssi}
\end{figure}

\textit{Carbon: }To provide some perspective, Fig.~\ref{f:cvssi} provides a linked view of the bulk carbon content of solar system bodies and the carbon carried by interstellar dust grains which represent a likely source term for the carbon of inner solar system solids.

The total abundance of carbon in the young sun was C/H\,$=2.88\times10^{-4}$ \citep{Lodders2003} and interstellar carbon grains carry about 50\% of this carbon. Within the solar nebula, accounting from comets Halley, 67P, and others suggests that this carbon was in refractory form \citep{Geiss87, Fomenkova99, Bergin15, Rubin19, Woodward21} in abundances close that seen in interstellar carbon grains.  The best accounting is from comet 67P \citep{Rubin19} and shows that a $\gtrsim5$-to-1 majority of elemental carbon in such objects is carried by refractories and not volatile ices.  Refractory carbon is found in meteoritic material, tracing the asteroid belt, but in amounts lower than that in comets by a factor of ${\gtrsim}10$, depending on the meteoritic class \citep{Alexander13, Bergin15}. In contrast, primitive meteorites are uniformly carbon-depleted as is the Earth and, likely, Venus.  The Earth's surface has significantly less carbon (see \S\,2.1), with the possibility of significant carbon reservoirs in the core.  Estimates of the Bulk Earth carbon content provide stringent upper limits that confirm Earth's carbon depletion at least to levels seen in meteoritic material \citep{Li21}.  Thus some aspects of reset (or loss) appear to be present in the solid state inventory.

In terms of refractory carriers, the dominant form for comets is believed to be similar to the macromolecular materials comprised of C, H, O, N in meteorites \citep{Alexander07, Fomenkova94}, along with lower abundances of aliphatic and aromatic compounds.  Meteoritic organics also have significant deuterium enrichments \citep{Alexander10}.  It is this material that has the most relevance for terrestrial planets as its sublimation temperature is estimated to be of order 500~K for nebular pressures \citep{Li21}.  

The most volatile material, carried only by the coldest bodies in the solar system, is CO and CO$_2$ ices \citet{Gail17}. The evidence from Arrokoth \citep{Grundyetal2020} and other KBOs, and from the CO-dominated comet C/2016\,R2 \citep{WierzchosWomack2018, McKayetal2019}, further establishes that both CO and CH$_{4}$ were significant volatile carbon reservoirs in the outer solar nebula. All three species sublimate at temperatures below ${\sim}70\,$K and should have been present in the gas  in the inner disk.
However, as discussed above the refractory inventory of the inner solar system also suggests that substantial refractory carbon must have been released to the gas at some stage, perhaps alongside the more volatile ices.


\textit{Nitrogen: }The story of nitrogen, to some extent, mirrors that of carbon as all inner solar system bodies (Earth, meteorites) are even more severely depleted in nitrogen when compared to carbon \citep{Bergin15, Marty20}.  However, fascinating new results have emerged since PPVI.
The total protosolar nitrogen abundance was N/H\,$=7.94\times10^{-5}$ \citep{Lodders2003}.  An absorption feature at $3.2\,\mu$m in \emph{Rosetta} spectroscopy of comet 67P has been attributed to ammonium salts, NH$_{4}^{+}$X$^{-}$ \citep{Quiricoetal2016, Pochetal2019}. These may constitute ${\leq}40\,$wt\% of the cometary surface material which contributes to the NIR spectrum \citep{Pochetal2020}. A more conservative estimate of $5\,$wt\% ammonium formate implies a cometary elemental nitrogen budget where ${\sim}1\,$\% is in volatiles (HCN, NH$_{3}$, etc.), ${\sim}47\,$\% in ammonium salts, and ${\sim}52\,$\% in refractory organics \citep{Pochetal2020}.
The Dawn mission detected ammoniated phyllosilicates on dwarf planet Ceres \citep{deSanctisetal2015}. Recent laboratory results support the feasibility of forming these phyllosilicates from NH$_{3}$ accreted at Ceres' formation \citep{Singhetal2021}. Forming these salts in the cometary ice required the presence of NH$_{3}$, which would need to be inherited from the protostellar core stage (see also Section\,5.1) or be produced from N$_{2}$ dissociation \citep{SchwarzBergin2014}. 

Nitrogen in the comae of Oort cloud comets is generally mostly in NH$_{3}$ and not N$_{2}$. The hypervolatile-rich comet C/2016\,R2 (PANSTARRS) was the first comet ever observed to have a high N$_{2}$ abundance \citep{WierzchosWomack2017, WierzchosWomack2018, Biveretal2018, McKayetal2019}. The relative role of N$_{2}$ in the spectroscopically observed elemental nitrogen budget was ${>}500$ times larger than that of NH$_{3}$ \citep{McKayetal2019}, likely indicating inheritance of N$_{2}$ as a dominant primordial nitrogen carrier in a relatively cold, shielded region of the outer protosolar nebula \citep{WierzchosWomack2018}.

If N$_{2}$ carried ${\gtrsim}99\,$\% of all volatile N in natal solar system material, the observation that it was largely processed into NH$_{3}$ seems to imply a relatively high cosmic-ray like ionizing particle flux in the comet-forming region.  However, nitrogen is present in refractory organics within meteoritic material and, at present, the relation between organic carriers and the nitrogen carried by salts is not clear.

\citet{Grewal21b} found evidence for distinct $^{15}$N/$^{14}$N ratios in NC vs. CC meteorites, implying a relatively refractory carrier of nitrogen (e.g., N-bearing complex organics or dust) was present already very early on in the solar system. The $^{15}$N/$^{14}$N ratio of the BSE falls between the NC and CC values, suggesting Earth's current nitrogen reservoir was accreted from both inner and outer solar system materials. The contribution of CC-like materials was found to fall between 30-60\% depending on the assumed total nitrogen budget of NC vs. CC bodies.

\textit{Hydrogen, oxygen, and water: }We treat the H, D, and O nuclei together because water is a volatile molecule whose presence is likely a crucial condition for habitability, while the D/H isotope ratio in water and other compounds is a useful diagnostic of the physico-chemical history of a given solar system sample. Our story of the deuterium-to-hydrogen ratio begins at the creation of time itself, with the baseline D/H ratio for solar system materials being set at the Big Bang \citep{Cookeetal2018} and remaining essentially unaltered as D/H\,$=(2.0\pm0.1)\times 10^{-5}$ in the local ISM \citep{Prodanovicetal2010}. Low-temperature chemistry can, however, lead to a highly significant elevation of D/H along certain chemical routes, such that HDO or D$_{2}$O abundances become comparable to that of H$_{2}$O at the $1\,$\% or even closer level. D/H ratios in various carriers (comets, meteorites) show modest  (factors of a few variation) diversity but in comparison to the elemental inventory are relatively flat \citep{Hartogh11, Cleeves14, Altwegg15} while gradients the oxygen isotopic composition are uncertain \citep{Altwegg15}.

The oxygen abundance in our sun at formation was O/H\,$=5.75\times10^{-4}$ \citep{Lodders2003}. In a solar composition gas, equilibrium condensation calculations place ${\approx}23\,$\% of oxygen atoms into refractory minerals, while volatile compounds such as H$_{2}$O carry the remaining ${\approx}77\,$\% \citep{Lodders2003}. Chemical kinetics, transport, and growth processes complicate attempts to directly link this to the refractory-to-volatile ratio in solar system bodies, however. From the empirical side, the actual ratio of refractories to volatiles in small icy bodies is difficult to pin down.  

The Jupiter-family comet 67P/Churyumov-Gerasimenko (67P) is thought to come from the scattered Kuiper Belt. During each passage close to the sun, it loses a significant surface layer, revealing pristine inner material. Determination of the dust-to-gas ratio in 67P vary from ${\lesssim}1$ to ${\gtrsim}6$ \citep{Choukrounetal2020} with a best estimate of $2.3^{+0.20}_{-0.16}$ based on the total mass loss \citep{ORourkeetal2020}. Volatile oxygen in 67P was almost entirely in H$_{2}$O ice. The other major oxygen carriers had a low relative abundance: CO$_{2}$/H$_{2}$O~$=(4.7\pm1.4)\%$, CO/H$_{2}$O~$=(3.1\pm0.9)\%$, and O$_{2}$/H$_{2}$O~$=(3.1\pm0.9)\%$ \citep{Bieleretal2015, Rubinetal2019}. The presence of free O$_{2}$ even at this level was unexpected and evidence points to its inclusion in the comet during assembly \citep{Bieleretal2015}. Chemical models also point to a pre-solar cloud origin for the O$_{2}$ molecules \citep{Taquetetal2016, Heritieretal2018}, consistent with a thermally unprocessed volatile reservoir in the comet-forming zone. An origin of O$_{2}$ in clathrates in a cooling protosolar disk (i.e., post-reset) has also been put forward \citep{Mousisetal2016}.

The high deuteration fraction of water in comet 67P \citep{Altweggetal2015} is likely primordial, based on results from protoplanetary disk ionization and chemistry models that show high D/H ratios can only be inherited from the cold, protostellar cloud \citep{Cleeves14}. In contrast to comet 67P, the cold classical Kuiper-belt object (KBO) 486958\,Arrokoth has remained in the icy outer solar system at 45 au, its primordial surface being rich in CH$_{3}$OH and (surprisingly) poor in H$_{2}$O \citep{Sternetal2019, Grundyetal2020}. While the former can be formed through hydrogenation of CO ice, the simultaneous depletion of H$_{2}$O also suggests radiolysis of water and CH$_{4}$ as the formation channel \citep{Grundyetal2020}.

 A central perspective is that oxygen can be carried to inner solar system bodies via silicates. However, given the generic abundance of water ice in outer solar system bodies, it is clear that the Earth and most meteoritic pre-cursors formed inside the nebular ice line and are hydrogen, i.e. water, poor \citep{Morbidelli00}.

\textit{Sulfur: }The sulfur abundance in the young sun was S/H\,$=1.89\times10^{-5}$ \citep{Lodders2003}. Based on meteoritic and cometary data, there is evidence for a dominant role for volatile sulfur (e.g., H$_{2}$S) in pristine inherited material, while sulfur in the thermally reset inner solar system reservoir is almost entirely in refractory carriers (e.g., FeS).

The simultaneous presence of a high abundance of volatile and refractory sulfur in comets testifies to a remarkable efficiency of transporting sublimated, re-condensed (or re-frozen) material into the comet-forming zone prior to comet formation. Earlier laboratory work on FeS formation demonstrated the feasibility of converting H$_{2}$S to FeS through interaction with exposed solid Fe surfaces at temperatures relevant to the early solar nebula \citep{Laurettaetal1996}. The \emph{Stardust} mission later found a high abundance of FeS in dust from 81P/Wild\,2 \citep{Westphaletal2009}. Radial mixing (the mechanism of which is as-yet unclear) may have transported this processed material outward to mix with pristine material in the comet-forming zone \citep{Westphaletal2009}.

Two recent measurements strengthen the case for sulfur having been volatile in inherited material at solar system formation. Firstly, \emph{Rosetta} measurements of volatiles in comet 67P/C-G reveal H$_{2}$S as the dominant outer solar system sulfur reservoir, at H$_{2}$S/H$_{2}$O\,${=}1.1\,$\% \citep{Calmonteetal2016, Rubinetal2019}. Atomic S was second at ${\approx}42\,$\% of H$_{2}$S, followed by SO2, SO, OCS, and CH$_{3}$SH at ratios ${\leq}4\,$\%. The evidence supports only a relative trace abundance of sulfur chains (S$_{\rm n}$; $n\in{[2,8]}$) in comets. Secondly, a previously unidentified absorption feature at $1.8\,\mu$m in the spectrum of the KBO 486958\,Arrokoth was found to be consistent with a sulfur-rich tholin-like organic residue which could have formed through photochemistry in H$_{2}$S-rich ice \citep{Mahjoubetal2021}.


\textit{Phosphorus: }The young sun had a total phosphorus abundance of P/H\,$=3.47\times10^{-7}$ \citep{Lodders2003}. The fractional abundance in dust from comet Halley was P/H\,$=2.92\times10^{-7}$ \citep[scaled from P/Mg using proto-solar Mg/H;][]{Schulzeetal1997}. This leaves little for volatiles, consistent with the low abundance of volatile P identified in the coma of comet 67P. Volatile phosphorus carried by PO was found in comet 67P at a total abundance P/O\,${\approx}10^{-4}$ \citep{Rubinetal2019}. Given a solar P/O\,${=}5\times10^{-4}$, the above results are consistent with most elemental phosphorus being contained in cometary dust. In CC meteorites, phosphorus is carried primarily in the oxidized mineral apatite. In enstatite chondrites, however, phosphorus is found mainly as the reduced mineral schreibersite. Of particular note, CI chondrites carry near solar abundance of phosphorous \citep{Wasson88} within the aforementioned refractory minerals.

\subsection{Summary}
In summary, the Earth's CHNOPS budget (Sect.~\ref{subsec:earth_chnops}) is the result of a complex and drawn out accretion of a variety of building blocks (Sect.~\ref{subsec:accreting_earth}), the ingredients of which have their origins in the solar nebula and beyond (Sect.~\ref{subsec:protosolarnebula}). Recognizing that the Solar System is just one possible outcome of the long and windy star \& planet formation process, we now set out to summarize important and recent observations of CHNOPS in diverse astronomical settings, interpreting these data in the context of the formation of (exo)planetary systems. The aim is to highlight diversity and variation, emerging trends, and gaps in our understanding of how CHNOPS elements arrive in terrestrial worlds.

\section{CHNOPS IN THE COSMOS}\label{sec:CHNOPS_COSMOS}

The CHNOPS elements have different nucleosynthetic origins in stars and thus vary in relative abundances and proportions as generations of stars are born and die and move throughout the Galaxy; this process has traditionally been known as Galactic chemical evolution (GCE), although more recently it has expanded to include a more complete picture of Galactic ``chemo-dynamical'' evolution.

\subsection{Variations in stellar CHNOPS}\label{sec:stellar_chnops}
While measuring the compositions of planets outside the solar system is a nascent field, studying the detailed compositions of stellar photospheres is a well-established cornerstone of astrophysics. Just like we used the Sun as a proxy for the Solar Nebula (Sect.~\ref{subsec:protosolarnebula}), stellar abundances can serve as a proxy method for measuring the chemical composition and variation in star and planet forming regions throughout the Galaxy. 

The Milky Way stars are often subdivided into components based on their kinematics and chemistry -- (1) the inner bulge, which has the highest density of stars, many with large inclinations relative to the disk plane (2) the diffuse halo, which contains only old metal-poor stars that have been suggested to subdivide into ``high-alpha'' and ``low-alpha'' abundance populations\footnote{\textbf{``Alpha'' refers here to elements whose most stable isotopes form via the $\alpha$-capture process in massive stars prior to Type II SNe explosions, e.g., Mg, Si, Ca, and Ti. O and S are often also considered alpha elements; see \cite{Nissen&Schuster2014} for a discussion of C.}} \citep[e.g.][]{Nissen2010, Bensby2014}, and (3) the disk, which is often divided into thin (higher metallicity) and thick (lower metallicity), where there is continuing star formation and a wide range of metallicites and ages of stars  \citep[e.g.,][]{Haywood2013}. Here ``metallicity'' can be broadly understood as the abundance of elements heavier than H, but it is often parameterized by the iron abundance, which increases with time as Type Ia supernovae (arising from long-lived white dwarfs accreting enough mass from a binary companion to trigger carbon fusion and runaway thermonuclear explosion) start to contribute more to the ISM.

\textit{Hydrogen} (along with helium) is the starting material from which the earliest stars were born, and originated not in stellar nucleosynthesis (as in the other CHNOPS elements) but soon after the birth of the universe; $\sim$380,000 years after the Big Bang the universe had cooled enough for neutral atoms to form. Hydrogen fusion is the main source of energy in the centers of stars on the main sequence, mainly through the proton-proton chain reaction (dominant in lower-mass/cooler core stars) and/or the CNO cycle (dominant in higher mass/hotter core stars). Given its overwhelming abundance relative to other elements ($\sim$74\% of baryonic matter), other elemental abundances are often quoted as relative to hydrogen, with the solar abundance measured/set at 10$^{12}$ H atoms (see discussion in \citealt{Lodders2019}).

\emph{Carbon} is produced via the triple-$\alpha$ process during the end stages of low- to intermediate-mass stars' evolution, and ejected into the interstellar medium (ISM) through either stellar winds (lower mass stars) or Type II (core-collapse) supernova explosions (more massive stars). [C/Fe]\footnote{This notation refers to the relative number density of carbon atoms in a star's photosphere relative to the amount of iron atoms, normalized to these values in the Sun, in log units: [C/Fe] = [C/H] - [Fe/H], where [X/H]=log(N$_{\rm{X}}$/N$_{\rm{H}}$) - log (N$_{\rm{X}}$/N$_{\rm{H}}$)$_{\rm{solar}}$. Thus the solar value in this notation is equal to zero.} is between $\sim$-0.15 and 0 dex for most thin disk stars, and rises from $\sim$0 dex to a plateau of $\sim$0.3 dex for thick disk and high-alpha halo stars; low-alpha halo stars (with [Fe/H] $\lesssim$-0.8 dex) show a range of [C/Fe] values between roughly -0.3 and 0.1 dex \citep[e.g.,][]{Nissen2014}.

\emph{Oxygen} is produced in hydrostatic burning in massive stars that then explode as Type II supernovae and inject oxygen into the ISM. The abundances of [O/Fe] show a tighter trend with [Fe/H] in thin disk stars, increasing from [O/Fe]$\sim$-0.2 dex around [Fe/H] of 0.3 dex to 0.2 dex around [Fe/H] of -0.4 dex, when thick disk and high-alpha halo stars take over to continue the upward trend to [O/Fe]$\sim$0.8 dex. The low-alpha halo stars follow a similar trend in [O/Fe] vs. [Fe/H] as the high-alpha halo stars, but offset to slightly lower [O/Fe] values. [C/Fe] and [O/Fe] are actually correlated and well-fit with a line of slope one, and thus the C/O ratio shows a tight, increasing trend from about 0.4 to about 0.8 dex as a function of [Fe/H], although importantly C/O does not actually appear to exceed 0.8 (see below) even at the highest [Fe/H] values \citep{Nissen2013,Teske2014,Brewer2016}.

\emph{Nitrogen} is produced in the CNO-cycle, which catalyzes H-burning in stars, with C and O decreasing as N increases. The abundances of nitrogen are harder to measure than C and O (which are already challenging) due to the lines being very weak, and thus less reliable. A handful of studies suggest that [N/Fe] is $\sim$constant across a wide [Fe/H] range ($\sim$-1.0 to 0.1 dex) \citep{Clegg1981,Laird1985,Carbon1987,Chiappini1999,Shi2002}, although precisely what that constant is is not clear due to systematic errors causing scatter. A roughly constant [N/Fe] suggests a late source of N (such as low and intermediate mass stars; \citealt{Prantzos2003,Chiappini2003}) is also needed to match the late source of Fe. 

\emph{Sulfur}, similar to oxygen, is an alpha-element produced in massive stars and injected into the ISM through Type II supernovae explosions. Around solar [Fe/H], [S/Fe] is slightly sub-solar, and at higher metallicities is appears to increase above solar by $\sim$0.2 dex. [S/Fe] then increases with decreasing [Fe/H] in both thin and thick disk stars to a plateau at a [S/Fe]$\sim$0.3 dex between -3.5$\leq$[Fe/H]$\leq$-0.1 dex, although with a scatter of about $\pm$0.2 dex \citep{Chen2002,Nissen2007,Spite2011,Matrozis2013,Duffau2017}.

\emph{Phosporous} is produced via neutron capture on silicon, a process thought to occur in the hydrostatic neon-burning shells of massive stars prior Type II supernovae explosions, and also during the explosion itself in carbon- and neon-burning layers \citep{Koo2013}. Similar to other alpha-elements, [P/Fe] increases as [Fe/H] decreases down to [Fe/H]$\sim$-0.5 dex, at which [P/Fe] plateaus around 0.3 dex; [P/Fe] is close to zero for solar metallicity, and perhaps increases with [Fe/H] above solar \citep[e.g.,][]{Caffau2007,Melendez2009,Caffau2011,Roederer2014}.

\subsection{Radionuclides}
In addition to the CHNOPS elements of focus, the chemical habitability of a planet will be influenced by the system's inventory of long-lived radionuclides (half lives $\gtrsim$700 Myr, e.g., $^{235}$U, $^{238}$U, $^{232}$Th, $^{40}$K), key to a planet's heat budget (Lichtenberg et al., in this volume) and short-lived ones (half lives $\lesssim$100 Myr, e.g., $^{26}$Al and $^{60}$Fe), which, if in sufficient abundance, can lead to differentiation and dehydration/devolatilization of planetesimals (Sect.~\ref{sec:planetesimal_heating}).

Uranium and thorium are produced via $r$-process nucleosynthesis, likely in neutron star merges \citep[e.g.,][]{Pian2017,Kasen2017}, while $^{40}$K is produced via oxygen-burning and the $s$-process in massive stars. 
$^{26}$Al can be injected into the ISM from core-collapse supernovae or from stellar winds emitted during post-Main-Sequence evolution of massive stars \citep{Gaidos2009,Gounelle2012,Young2014}. Much of our knowledge about $^{26}$Al actually comes from observations of the 1809 keV gamma rays produced when it decays \citep{Reiter2020}, which support the idea that $^{26}$Al is produced by high-mass stars, with up to 50\% coming from pre-supernova mass loss, especially from Wolf-Rayet stars \citep{Crowther2007}. Iron-60 is likely produced in the helium- and/or carbon-burning shells of massive stars via neutron capture on preexisting stable iron isotopes (see \citealt{Wang2020} for further details).

The levels of short-lived radionuclides (particularly $^{26}$Al and $^{60}$Fe) found in meteorites are greater than that expected from steady-state production in the Galaxy \citep[although see][]{TangDauphas2012}, indicating there was likely a local source of enhancement within $\sim$1 Myr and $\sim$0.2 pc of the birth of the Solar System \citep{Harper1996,Wasserburg1996,MeyerClayton2000,Adams2010}. The enhanced short-lived radionuclides could have come from (1) a supernova explosion or mass loss from a ``super''-asymptotic giant branch star near the Sun's protoplanetary disk \citep{Chevalier2000,Ouellette2007,Lugaro2012}, although based on simulations of stellar cluster evolution this seems unlikely \citep{Parker2014}, (2) a triggered cloud collapse with accompanying direct injection of the radionuclides from a single supernova \citep{Cameron1977,FosterBoss1997,Boss2015}, or (3) a more sequential process in which the Sun formed in an inter-generational giant molecular cloud enriched first in $^{60}$Fe and then $^{26}$Al, the latter perhaps from winds from Wolf-Rayet stars \citep{Arnould2006,Gaidos2009,Young2014,Dwarkadas2017}. Whether $^{60}$Fe is mildly or strongly under-produced vs. $^{26}$Al, and thus how decoupled the sources of $^{26}$Al and $^{60}$Fe were in the early Solar System, is still a topic of active study \citep[e.g.,][]{NittlerCiesla2016}. See Desch et al. (in this volume) for a more comprehensive review of short-lived radionuclides in the context of the early Solar System.

All of these radioactive elements are expected to be mixed in the ISM from shear forces in the differentially-rotating Milky Way disk, on the order of 250 Myr \citep{Frank2014}, meaning the short-lived nuclides could experience inhomogenous dispersal within their half-lives (and indeed may be differently distributed, see \citealt{Wang2020}), but longer lived nuclides are likely more homogenously dispersed at a given Galactic radius \citep{Huss2009}. \citet{Reiter2020} concluded that the ISM is regularly replenished with high levels of $^{26}$Al that can be sustained for millions of years, and if $\sim50$\% of stars are born in high-mass star forming regions \citep[e.g.][]{Dukes2012}, then as many as $\sim$25\% of systems may be enriched in $^{26}$Al at a level similar to the early Solar System. However, it is important to note that the large-scale emission observed from $^{26}$Al is irregular \citep{Wang2009}, indicative of a nonuniform distribution of massive stars, and even if a giant molecular cloud is in the vicinity of a massive star association, it does not guarantee $^{26}$Al will find its way into the cold clumps of star formation (see \citealt{Lugaro2018} for an extended discussion). \citet{Lichtenberg2016} modeled this process with $N$-body simulations of large (10$^{3}$-10$^{4}$) clusters of stars with a range of initial conditions, and found a wide range of resulting short-lived radionuclide enrichment factors, with ${\sim}10-30$\% of systems enriched commonly at Solar System levels, but also many systems with negligible or zero short-lived radionuclide abundances.

\section{LESSONS FROM EXO-PLANETS AND THEIR HOST STARS}\label{sec:exoplanet_hosts}
Before continuing our journey of CHNOPS from their cosmic origins to molecular clouds and protoplanetery systems (Sect.~\ref{sec:3clouddisk}), we take a moment to highlight important lessons from studies of exoplanets and their host stars. While measuring detailed CHNOPS budgets on/in individual extra-solar terrestrial worlds is currently out of reach, it has become possible to study emerging trends between the occurrence rates of different planet types and host star properties. We choose here to highlight two selected host star characteristics: metallicity ([Fe/H]) and CHNOPS budget. For more in-depth assessments of the lessons from \textit{Kepler} and emerging trends in exoplanetary architectures, we point the interested reader to Lissauer et al., and Weiss et al., (this volume).

\subsection{Host star metallicity}\label{sec:exo_host_metallicity}

Giant planets (at least in relatively short orbital periods) are observed more frequently around metal-rich stars \citep[e.g.,][]{Santos2004,Fischer2005,Johnson2010,Mortier2013}, while smaller planets appear to form around stars with a wider variety of metallicity \citep[e.g.,][]{Gonzalez1997,Sousa2011,Buchhave2012,Buchhave2014,Wang2015,Petigura2018}. The predicted fraction of stars with giant planets is often parameterized as a function of metallicity using a power law, $f$([Fe/H]) $\propto 10^{\beta \rm{[Fe/H]}}$, and previous studies have found $\beta$ for hot ($P<10$ day period) Jupiters to range from 0.71$^{+0.56}_{-0.34}$ \citep[][although note this study was not sensitive to the absolute occurrence rate]{Osborn2020} to 3.4$^{+0.9}_{-0.8}$ \citep{Petigura2018}. The index $\beta$ then decreases for smaller planets, e.g., Petigura et al. found $1.6\pm0.3$ for hot sub-Neptune planets and $0.6\pm0.2$ for hot super-Earths in the \textit{Kepler} sample of FGK dwarf stars between [Fe/H] $= $-0.4 and 0.4 dex. Interestingly, the metallicity dependence measured by Petigura et al. appears to weaken or flatten out for longer period planets of all sizes \citep[consistent with e.g.,][]{Mulders2016,Buchhave2018}. However, how metallicity impacts the formation and resulting compositions of different planets is still somewhat qualitative and deserves continued study as more detailed exoplanet data are measured and larger spectroscopic surveys of stars are conducted. For example, it is curious that \citet{Teske2019} did \textit{not} find a clear correlation between stellar metallicity and planet residual metallicity (the relative amount of metal versus that expected from the planet mass alone) in a sample of non-inflated giant exoplanets, which conflicts with common predictions from formation models that a feeding zone of a forming giant planet with twice as much metal should produce a giant planet with $\sim$twice as much metals.

Several RV searches for planets around metal-poor ($-2.0\leq$[Fe/H]$\leq$-0.6 dex) stars resulted in a prediction for the average frequency of giant planets (hot Jupiters with periods $<$10 days) $\lesssim$2\%, versus their frequency at higher metallicities of $\sim$3-8\% \citep{Sozzetti2009,Mortier2012}. The recent work of \citet{Boley2021} made use of the boon of TESS observations of halo stars to search for transiting hot Jupiters in the same metal-poor regime, resulting in an mean 1-$\sigma$ upper limit of 0.18\% for 0.8-2 R$_{J}$ planets orbiting with periods between 0.5-10 days. But due to the technical challenges of detecting smaller planets around metal-poor stars, which have fewer absorption lines and are generally farther away and thus fainter, there is currently no strong observational constraints on, ``How low can you go?'' for small planet host star metallicity. The lowest-metallicity stars known to host a planet with $M_p \leq 10 M_{\oplus}$ or $R_p \leq 4 R_{\oplus}$ have Fe/H]$=$ -0.89 dex (Kapteyn's star; \citealt{Anglada-Escude2014}) and [Fe/H]$=$ -0.65$\pm$0.10 dex (K2-180, \citealt{Korth2019}), respectively. Interestingly, observations of iron-poor planet-hosting stars (mostly larger planets) show that the host stars exhibit preferential enhancement in alpha-element abundances, indicating that perhaps other heavy elements may compensate for solid planetary building blocks when iron is lacking \citep{Brugamyer2011,Adibekyan2012a,Adibekyan2012b,Adibekyan2015}.

\subsection{Host star CHNOPS abundances}\label{sec:exo_host_chnops}
With increasing interest in distinguishing populations of exoplanets and their potential to host life, the ratio of carbon to oxygen (C/O) in exoplanet host stars has also been the focus of detailed study. The ratio of carbon to oxygen (along with that of magnesium to silicon) significantly influences the bulk minerology planets, such that a C/O ratio $>0.8$ will produce carbon-rich planets dominated by graphite, silicon carbide, and titanium carbide \citep[e.g.,][]{Gaidos2000,Kuchner2005} and likely geodynamically inactive \citep{Unterborn2014}.

As noted above, despite significant variations in [C/Fe] and [O/Fe] with [Fe/H], these trends do not depart from each other very much and thus the C/O ratio spans only about 0.4 to 0.7 (between -0.4 and 0.5 dex in [Fe/H]) in high-precision chemical abundance studies, with no difference between stars known and not known to host planets \citep{Teske2014,Nissen2014}. The larger sample study (849 Sunlike RV planet search stars) of \citet{Brewer2016} showed a slightly wider distribution of C/O across roughly the same [Fe/H] (see their Figure 1), with a median C/O of 0.47 and only four stars even approaching C/O$=$0.8 (the maximum stellar C/O in their sample is 0.66$\pm$0.068); indeed, they suggest the Sun at C/O$=$0.55 is slightly carbon-rich. Other studies have found similar C/O spreads and the lack of a difference between the C/O ratio in stars known vs. not known to host planets \citep[e.g.,][]{Suarez-Andres2017,Suarez-Andres2018, Clark2021}, but it is important to note that these previous studies of C/O in planet host stars were dominated by stars hosting planets that are somewhat or much larger than what we would consider terrestrial. While \citet{Brewer2018} derived [C/H] and [O/H] ratios for $\sim$1000 \textit{Kepler} stars, many of which are known to host smaller planets, they do not report on differences between the distribution of C/O in different subpopulations of host stars.

There have been fewer studies of the variation in abundances of N, P, and S in planet host stars. \citet{Ecuvillon2004} presented a spectroscopic analysis of nitrogen in 66/25 stars with/without known planets, combining abundances derived from both the near-UV NH band and the near-IR N I line, and found no significant difference in the behavior of [N/Fe] vs. [Fe/H] in planet hosts. More recent studies have confirmed these results \citep{daSilva2015,Suarez-Andres2016}. Due to the challenges in measuring P abundances in general (see above), there are only $\sim$a dozen exoplanet host stars with P measured along with another element (Hypatia Catalog, \citealt{Hinkel2014}), so it is not possible to draw conclusions about the behavior of P in stars known to vs. not known to host planets, other than there is not obviously larger or smaller scatter between the two populations in most ratios (P/Si appears to show some high outliers that are planet hosting stars). For sulfur, the recent study of \citet{Costa_Silva2020} found no distinctive trends or differences between the behavior of [S/Fe] vs. [Fe/H] in planet host stars (110 Jovian-type planets, 24 Neptune- or super-Earth type planets) versus stars not known to host planets (585), although below [Fe/H] of -0.3 dex their sample suffers from a small-number statistics for the planet host stars.

\section{MOLECULAR CLOUDS AND THE FORMATION OF PROTOPLANETARY DISKS}\label{sec:3clouddisk}

The interstellar medium consists of phases which cover multiple orders of magnitude in density, temperature, and radiation field intensity. The CHNOPS elements aside from hydrogen are made in stars (Section\,\ref{sec:CHNOPS_COSMOS}), released into the diffuse ISM where they are mostly in atomic form, and may later get incorporated in molecular clouds. 

Dense regions of molecular clouds can become gravitationally bound pre-stellar cores, which can collapse in the protostellar core stage, forming a central protostar with an accretion disk (re-)supplied by the collapsing envelope. Up to the sublimation temperature, heating of ices can lead to increasing chemical complexity, including complex organics which sublimate at higher temperatures \citep{oberg2009d, Boogert2015}. Most of the new star's mass is built up within 0.5\,Myr \citep{machida2011, Hartmann16}. The disk starts out small and hot, but quickly grows and is thought to reach its largest angular extent and highest mass in the still warm Class\,0/I stage \citep{hueso2005}. ALMA observations confirms that Class\,0 disks are warm \citep{vanthoff20}. ``Time zero'', corresponding to CAI condensation in the innermost solar system, is thought to happen somewhere in the Class\,0/I stage. After this the disk loses angular momentum and mass due to viscous evolution and winds in the Class\,II stage. Viscous heating loses its importance and stellar irradiation largely dictates the temperature for most of the disk's lifetime.

Millimeter grains in most Class\,0 disks remain confined to ${\approx}60\,$au \citep{maury2019} and some have centimeter grains extending out to $12\,$au \citep{segura-cox2018}. Class\,II disks are both larger and more diverse in size \citep{NajitaBergin2018}. Dust growth starts very early, with growing grains increasingly settling on the midplane and drifting inwards due to gas drag. The surface-to-midplane vertical separation of irradiation heating and dust cooling result in a shielded midplane containing most of the mass, sandwiched between warmer molecule-rich layers a few scale heights from the midplane and a third, PDR-like layer which envelops the whole disk \citep[e.g.,][]{Aikawa2002,Henning2013}. Moving outwards along the midplane, the decreasing temperature gives rise to a sequence of snow- or ice-lines which may play a key role in deciding the CHNOPS content of any planets that form \citep{Oberg2011}. In the youngest, hottest disks, \ce{H2O} is only frozen out at radii ${\approx}10{-}100\,$au \citep{vantHoff2021} and CO is entirely in the gas. As these disks cool and radially spread, the snowlines of all ice and organic species move closer towards the star and CO can freeze out in the outer disk for the first time \citep[e.g. Class I, ][]{vanthoff20}. Eventually an N$_2$ snowline develops as well, as implied by gas-phase nitrogen depletion in mature disks \citep[e.g. TW Hya, ][]{vanthoff2017, hily-blant2019, mcclure2020}.

Infrared observations of protostars reveal pure \ce{CO2} ice \citep{pontoppidan2008}, requiring the complete sublimation of CO ice in most of the protostellar envelope which may in turn point to episodic accretion luminosity bursts \citep{kim2012, visser2015, Hartmann16}. Such bursts would also impact other CHNOPS ice species, pushing snowlines outwards by tens of au \citep[e.g.,][]{Taquet2016}. Spectra showing crystalline \ce{CO2} signatures in some protostellar cores furthermore suggest repeated sublimation/condensation events of ices up to $70\,$K \citep{poteet2013}. ``Relic snowlines'' consistent with past outbursts have also been seen for \ce{H2O} \citep{vanthoff2018}. Some ices such as \ce{H2O} may survive these events in large parts of the envelope, contributing a relatively pristine material reservoir for disks \citep{visser2009}. The D/H ratio of Earth's water is also consistent with a significant inherited \ce{H2O} ice reservoir \citep{Cleeves14, Furuya2017}. An example of a system currently in outburst is V883 Ori, with an apparent \ce{H2O} snowline at $42\,$au based on dust emission \citep{Banzatti2015, cieza2016} or even ${\approx}100\,$au based on HCO$^{+}$ \citep{Leemker2021}. Such objects are excellent laboratories for gas-phase spectroscopy of evaporated ices, which for V883\,Ori show a comet-like COM inventory \citep{lee2019}.

The small angular size of disks in combination with the high optical depths and efficient freeze-out of volatiles in the outer disk make observing molecules challenging, and only about two dozen of the ${>}200$ molecules identified in the ISM and protostars have been found in protoplanetary disks \citep{McGuire2018}. At IR wavelengths, vibrational transitions of gas-phase \ce{H2O} and \ce{CO2} are seen in emission, allowing for an accounting of CNO-bearing molecules in the surface layers of the inner few au \citep{Carr2008,Najita2013,Pontoppidan2014}. In rare cases, ro-vibrational absorption features allow an accounting of otherwise nearly invisible species such as \ce{CH4} \citep{GibbHorne2013}. Molecules with a permanent dipole moment (e.g., CO, HCN, or \ce{CH3OH}) often have rotational transitions at (sub)millimeter wavelengths that can be seen with (sub-)millimetre telescopes such as ALMA \citep{Oberg2015,Walsh2016,Booth2021b}. Figure \ref{fig:as209} illustrates the spatial scales probed by IR and mm-wavelength observations for the disk of AS 209. Finally, in disks with favourable near-edge-on inclinations, ice species may be seen in absorption, allowing for the characterisation of their abundances as well as ice properties along selected sightlines \citep{pontoppidan2005,terada2007, Ballering2021}. 

\subsection{Elemental Accounting from the ISM to Protoplanetary Disks}

We next review the budget of each element on its journey from the ISM through to late-stage protoplanetary disks, forming a backdrop for a closer look at planet formation which follows in Section\,\ref{sec:planet_formation}.

\textit{Oxygen: }The depletion of oxygen in the diffuse atomic ISM is consistent with its depletion into silicate minerals. However, with increasing ISM density, oxygen depletes from the gas more rapidly than its partner elements in silicates, suggesting the existence of an additional significant (solid?) oxygen reservoir of unclear nature \citep{Jenkins2009}. Refractory organics are a likely candidate \citep{Whittet2010}. Based on the elemental budget, viable outcomes O and CO accretion, and spectroscopic constraints, it has been suggested that the missing oxygen reservoir is an organic carbonate, which would also tie up ${\sim}12$ to $19\,$\% of all elemental carbon \citep{JonesYsard2019}, but the latest oxygen budget reviews do not favour a significant refractory organic component for O \citep{vanDishoeck2021}. Focussing purely on the ices in dense clouds and molecular clouds \citep{Boogert2015}, oxygen appears first as amorphous H$_{2}$O ice. With the formation of CO ice, CH$_{3}$OH ice also forms. The ice budget in protostellar cores, where disk formation is thought to start, accounts for ${\approx}26$ to $60\,$\% of total elemental O as H$_{2}$O, CO, and CO$_{2}$. In the outermost regions of the late-stage protoplanetary disk HD\,142527, around $80\,$\% of elemental O is present as crystalline \ce{H2O} \citep{Min2016}.

\textit{Carbon: }The disposition of elemental carbon in interstellar space is such that roughly 50\% is thought to be in the form of refractory carbonaceous grains formed in asymptotic giant branch stars and 50\% in the gas, primarily as CO in diffuse and dense clouds \citep{Cardellietal1996, Jenkins2009, Parvathietal2012, Mishra15}. New results suggest that polycyclic aromatic hydrocarbons can form in cold (T\,$\sim$10~K), dense filaments \citep{McGuire21}. Thus there is the potential for large molecule/small carbon grain formation as part of the star formation process, though the abundance of the precursors is very low \citep{McGuire2018}. In protostellar cores, spectroscopically identified ices account for $14$ to $27\,$\% of C \citep{Boogert2015}. This does not include PAHs.

\textit{Nitrogen: }In the diffuse ISM, nitrogen is present as an atomic gas which, unlike carbon, shows no statistically significant depletion with increasing extinction \citep{Jenkins2009}. The $20\%$ level of depletion observed at $1\sigma$ significance may be related to N incorporated in carbonaceous nanoparticles \citep{Jones2016a}, a form of CHON material. In star-forming molecular clouds, observational estimates of the molecular nitrogen to ammonia ratio yield N$_{2}$/NH$_{3}=(170\pm100)$ \citep{Womacketal1992}. This appears broadly consistent with the value of ${\gtrsim}500$ estimated for the coma of the hyper-volatile rich comet C/2016\,R2, perhaps suggesting a dominant role for N$_{2}$ at the time of protostellar core formation. In such cores, ices carry ${\approx}11$ to $30\,$\% of total elemental N as NH$_{3}$, NH$_{4}^{+}$, and OCN$^{-}$, while the role of N$_{2}$ is empirically poorly constrained \citep{Boogert2015}. If N is released from \ce{N2} by He$^{+}$, both gas-phase interactions (with H$_{2}$) or grain surface processes (H) can convert it into \ce{NH3} \citep[e.g.,][]{Fedoseev2015}. The possible presence of a significant fraction of total N as ammonia and ammonium salts is consistent with the solar system findings described in Section\,\ref{subsec:protosolarnebula}. In the IM\,Lup protoplanetary disk, \citet{Cleeves2018} found no evidence of gas-phase nitrogen depletion, consistent with the directly identified high fractional abundance \ce{NH3}/\ce{H2O}\,$=7$ to $84\,$\% in TW\,Hya \citep{Salinas2016}. Finally, inside ${\approx}10\,$au in three disks, \citet{Pontoppidan2019} found little \ce{NH3} which may signify an efficient conversion back into \ce{N2}.

\begin{figure}[!ht]
\centering
\includegraphics[clip=,width=1.0\linewidth]{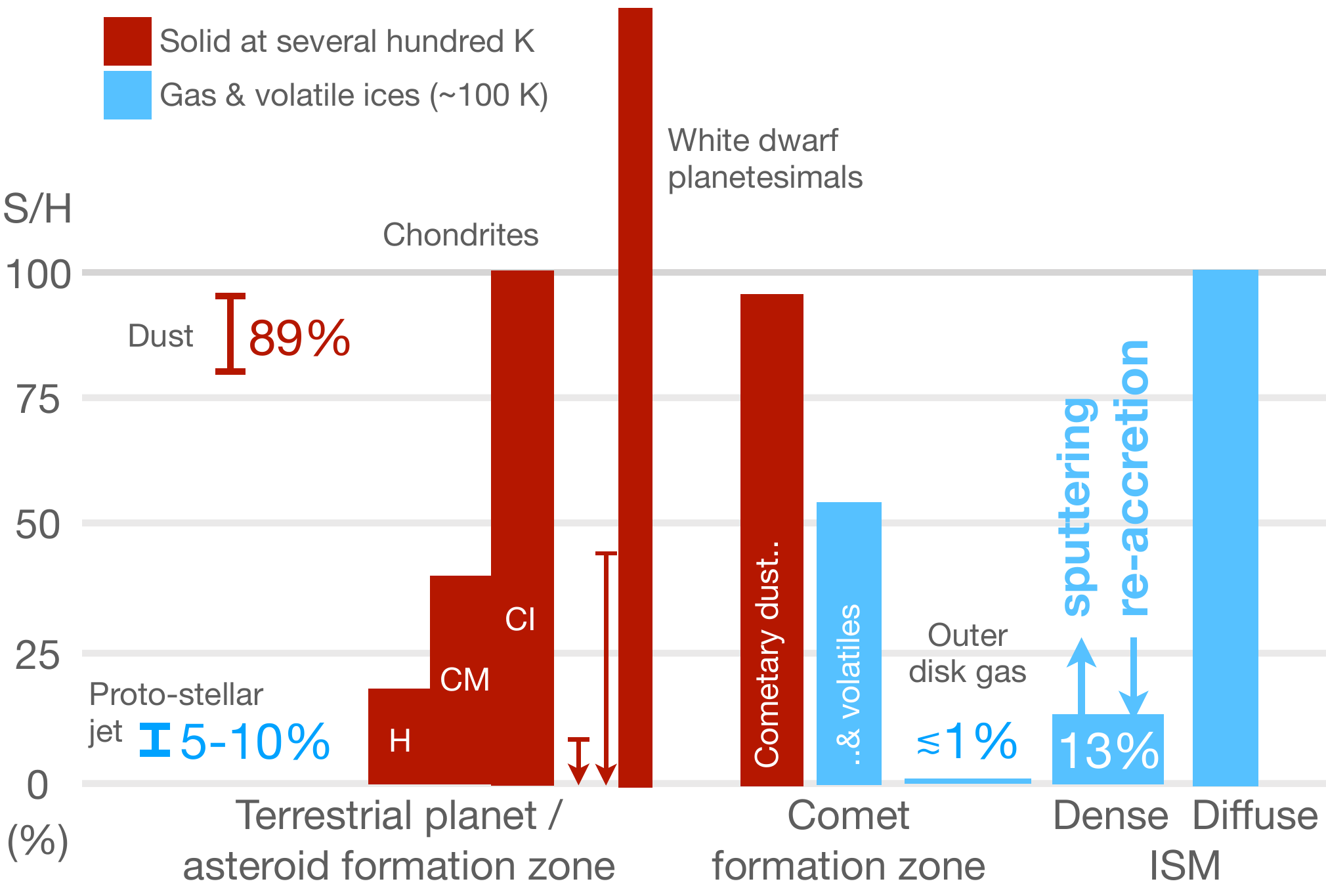}
\caption{The fraction of total sulfur locked in volatile (blue, $T_{\rm sub}{\lesssim}100\,$K) or refractory (red) carriers in the inner and outer disk, and in their natal ISM material. Values are normalised to solar, capped arrows are upper limits. Figure adapted from \citet{Kama19}.}
\label{fig:sulfurbar}
\end{figure}

\textit{Sulfur: }Sulfur is released by AGB stars and Type\,II supernovae in the form of volatile species such as SO \citep{Matsuuraetal2017, Danilovichetal2016, Danilovichetal2017, Danilovichetal2018} and as refractory minerals such as MgS \citep{Honyetal2002, Lombaertetal2012}, though the significance of the solid-state spectral feature underlying the MgS identification is disputed \citep[e.g.,][]{Volketal2020} and equilibrium condensation calculations suggest FeS is the dominant refractory form. In the diffuse atomic ISM, sulfur is entirely ``reset'' into an atomic gas \citep{Jenkins2009}. This is likely due to sputtering of sulfide minerals by high-energy particles \citep{Kelleretal2010}, though there is evidence that at least some FeS grains survived their interstellar journey, to be incorporated in meteorites as prestellar grains \citep{Haenecouretal2016}. With increasing column density, sulfur depletes from interstellar gas into a solid form of unclear nature. In protostellar cores, only $5\%$ of total elemental S has so far been recovered in ices (H$_{2}$S, OCS, and SO$_{2}$) \citep{Boogert2015}. In protoplanetary disks, ${\approx}89\,$\% of total S is accounted for by a fairly refractory species, likely FeS \citep{Kama19}, while gas-phase volatiles (\ce{H2S}, CS, SO) carry ${\lesssim}1\,$\% \citep{Dutrey1997, Fuente2010, Dutrey2011, PachecoVazquez2016, Semenov2018, ABooth2018}. The sulfur budget for planet formation is summarised in Fig.\,\ref{fig:sulfurbar}.

\textit{Phosphorus: }All elemental phosphorus is in atomic gas form in the diffuse ISM and it depletes rapidly towards higher densities \citep{Jenkins2009}. Very little is known empirically about the reservoirs phosphorus depletes into as it enters molecular clouds. Gas-phase PO and PN have been found to carry only $0.05$ to $0.5\,$\% of elemental P in star-forming regions \citep[e.g.,][]{Ziurys1987, Bergneretal2019}, while P-bearing ices themselves remain entirely undetected. Desorbed ices and sputtered refractory P are both potential sources for the gas-phase PN and PO \citep{Mininnietal2018}. It seems likely that nearly all P is locked in refractories, as it is in the solar system (Section\,\ref{subsec:protosolarnebula}).

\subsection{Disks: the context for planet formation}

As a snapshot of the current observational perspective in Fig.~\ref{fig:as209} which presents molecular emission images with $\sim$15 au resolution towards the AS 209 protoplanetary disk from the ALMA large program Molecules At Planet-forming Scales or MAPS \citep{Oberg21, Czekala21, Law21}.  The dust emission image (top left) exhibits significant sub-structure (gaps and rings) which are associated with one or more hidden planets \citep{Zhang2018}. In contrast, the molecular emission, exhibits evident sub-structure, but seemingly not concurrent with the dust \citep{Law21}.  This complex structure hints at a chemistry that is linked to the dust evolution but is also influenced by other factors such as the enhanced UV penetration as a result of dust growth and settling and elevated C/O ratios present on the disk surface \citep{Alarcon21, Bosman21, Guzman21, Ilee21}. ALMA images of molecular emission generally resolve into tens of au and do not readily probe the terrestrial planet forming zone ($0.1{-}3~\mathrm{au}$).  Molecular spectra at mid-infrared wavelengths can access this gas (and dust).  In AS209 the Spitzer spectrum exhibits emission lines of water, CO$_2$, HCN, and C$_2$H$_2$.  With the launch of JWST, and its anticipated science return there is a bright future to try to link the composition seen in ALMA images tracing the giant-planet forming zone (in our solar system) with that probed by JWST tracing gas coincident with the birth place of terrestrial planets.

Illuminating the temperature structure and cooling sequence in both young and old disks is of critical importance to planet formation, as evidence from the Solar System suggests that planetesimal formation started within 0.25 Myr after CAI formation and continued through $\sim$4 Myr \citep{NittlerCiesla2016}. One might then expect planetesimals formed in the early, warm disk phase to be poorer in CHNOPS than planetesimals formed at the colder, later stages. However, the exact CHNOPS composition of planetesimals also depends on various formation and disk dynamical processes, as we discuss next.

\begin{figure*}[t]
\centering
\includegraphics[clip=,width=1.\textwidth]{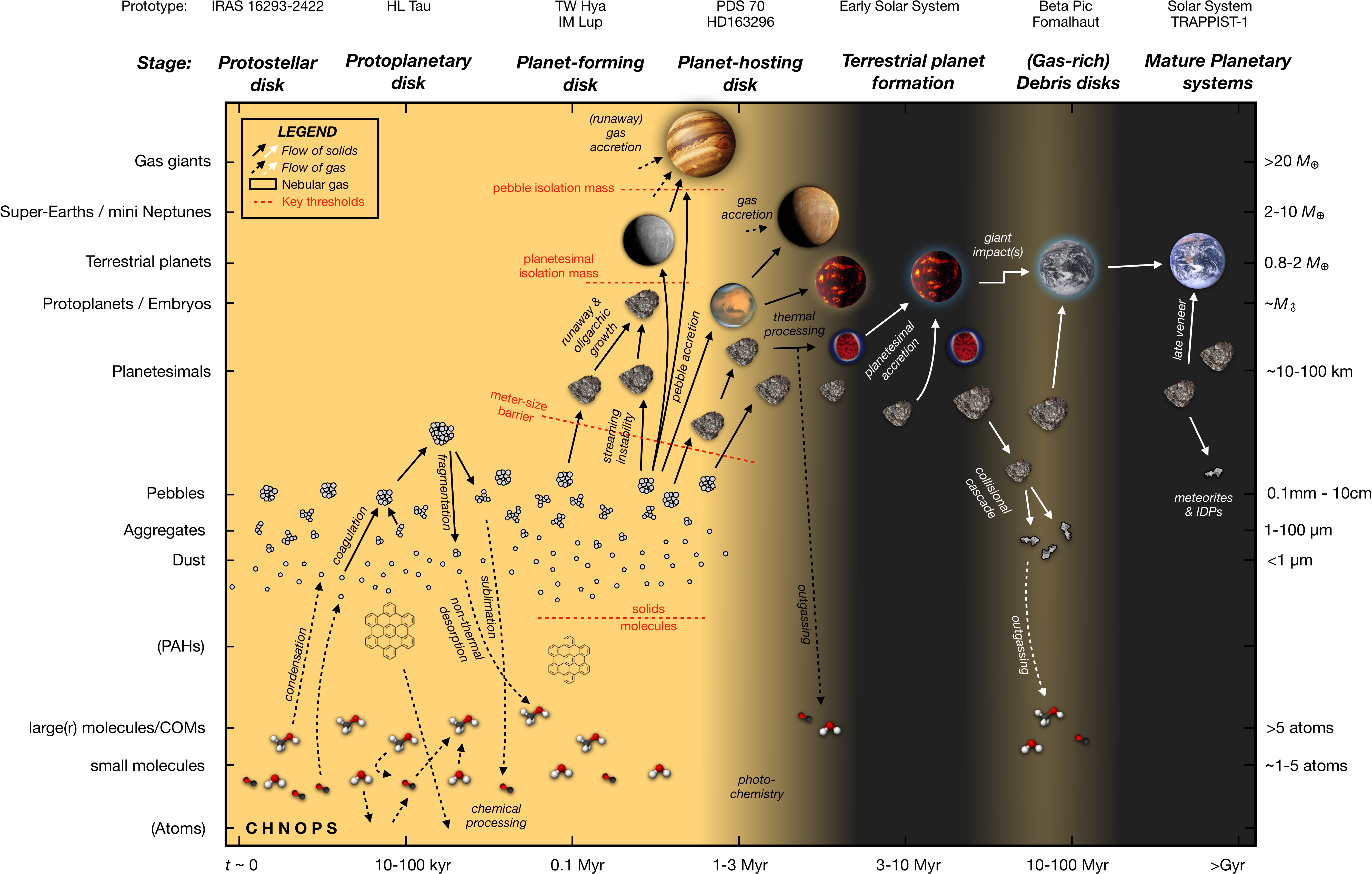}
\caption{Modern planet formation theory provides an unbroken chain of processes connecting nebular dust and gas (bottom left quadrant), through pebbles and planetesimals, all the way to giant planets, terrestrial planets, and everything in between. The picture is highly dynamic, with large-scale/disk-wide mixing and transport processes (not shown here) pervasive (and sometimes essential) at every mass/size scale. With this roadmap in hand, we can attempt to follow life-essential CHNOPS elements from the inter-stellar medium to mature and diverse worlds.}\label{fig:planetformation_cartoon}
\end{figure*}

\section{CHNOPS DURING PLANET FORMATION}\label{sec:planet_formation}

\subsection{Early stages of planet formation}\label{sec:dust_pebble_planetesimal}

\begin{figure*}[t]
\centering
\includegraphics[clip=,width=1.\textwidth]{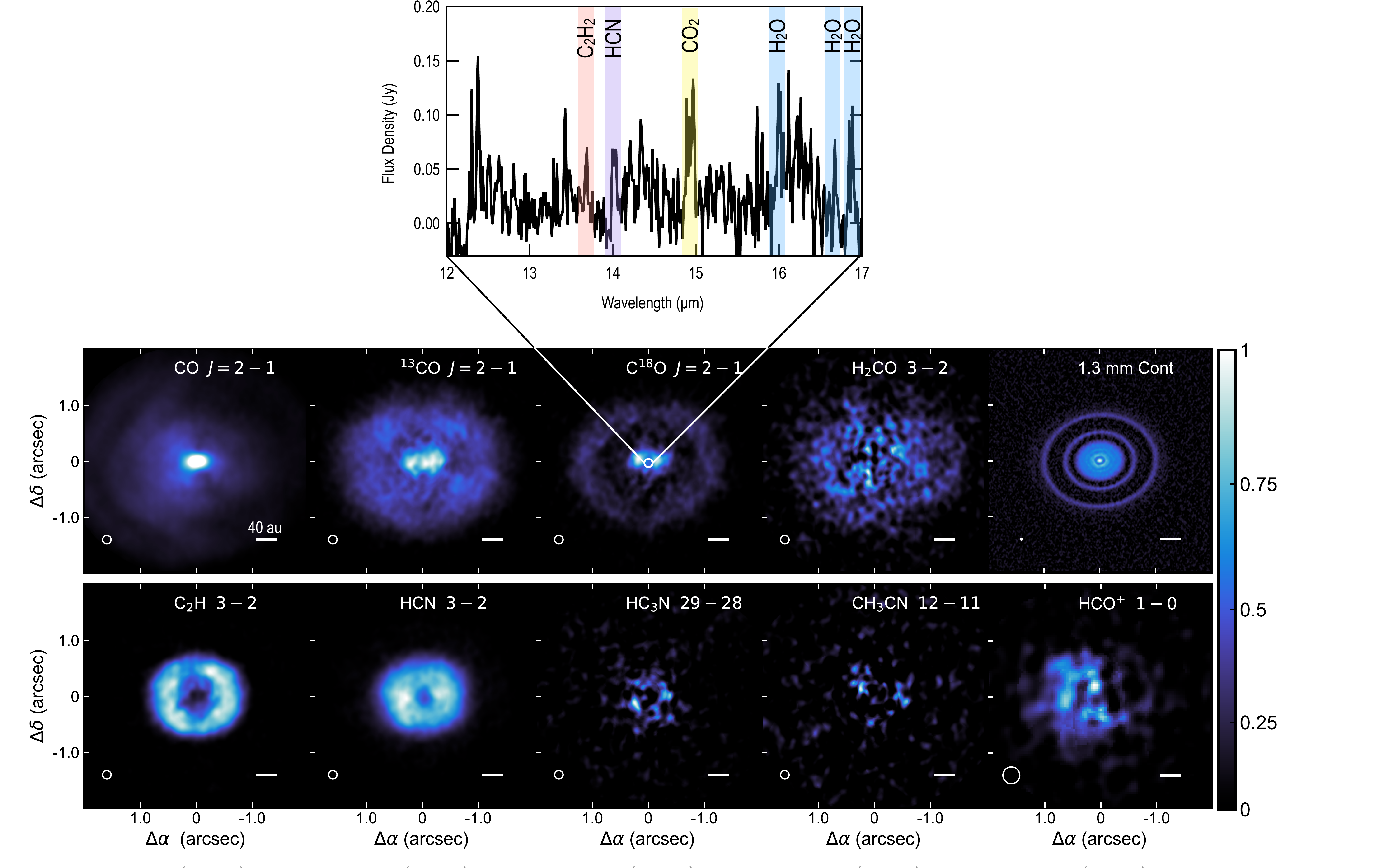}
\caption{
Images of the AS209 obtained as part of the Molecules at Planet-forming Scales (MAPS) ALMA 
large program \citep{Oberg21}.   Molecular emission images are from \citet{Czekala21} and \citet{Law21}.
Dust emission image is from the ALMA large program The Disk Substructures at High Angular Resolution Project or DSHARP \citep{Andrews18, Huang18}.    All emission images are normalized to the peak with $^{12}$CO
having a different normalization in order to enhance the weak extended emission.    Spitzer spectrum
of AS209 is shown on top with particular features identified.   This emission arises from the inner few
au of the disk as illustrated.   Spitzer spectrum kindly provided by C. Salyk.}\label{fig:as209}
\end{figure*}

\subsubsection{Chemical processing and mixing}

Variations in gas density, temperature, and the radiation field found in protoplanetary disks give rise to a rich combination of grain-surface and gas-phase chemistry on a variety of time-scales \citep[][and references there-in]{vanDishoeck2020,ObergBergin2021}. Moreover, advection and diffusion of molecules and (small) dust grains complicates the chemistry, as timescales for mixing can be comparable to those for chemical processing of materials \citep{Semenov2011}. Small grains, when mixed up to the surface, can be exposed to high temperatures and intense UV radiation, resulting in processing or loss of their ice mantles \citep{Ciesla2012,Bergner2021}, or, in the inner regions, even oxidation and photolysis of more refractory carbon carriers \citep{Lee2010,Anderson2017,Klarmann2018,Bosman2021}. Condensates from the inner nebula can also be transported outward, through viscous spreading \citep{Dullemond2006,NajitaBergin2018}, meridional flows \citep{Ciesla2007,Desch2018}, or by being picked up and flung out by disk winds \citep{Shu1996,Giacalone2019}, though winds may not be very effective at re-inserting swept-up grains into the outer disk \citep{BoothClarke2021}. In the Solar System evidence for (early) outward transport can be found in comets and CC meteorites \citep[][and references therein]{NittlerCiesla2016}. Signatures of a similar movement of solids in protoplanetary disks come from observations of crystalline silicate grains \citep[][]{vanBoekel2004,Apai2005}. However, as grains coagulate and grow larger, their dynamics change considerably and instead, in the absence of pressure bumps (see Sect.~\ref{sec:disk_structures}), inward radial drift along the midplane is expected to dominate \citep{Weidenschilling77,Misener2019}.

\subsubsection{CHNOPS and the meter-size barrier}\label{sec:meter_size_barrier}
Locally, microsopic dust particles collide and stick in collisions to form larger dust aggregates. This pairwise growth stops when aggregates reach the bouncing or fragmentation barrier (when collisions no longer result in sticking), or when radial drift removes aggregates on timescales that are shorter than their coagulation timescales \citep[e.g.,][]{Birnstiel2016}. Collectively, these bottlenecks are often referred to as the "meter-size barrier" (see Fig. \ref{fig:planetformation_cartoon}). CHNOPS elements and their partitioning between gas, ice, and solid phases influence this picture in several ways.

First, the condensation of major ices (water, CO, CO$_2$) increases the local mass surface density of solids, speeding up dust coagulation shifting the radial drift barrier. Typical (initial) values of the total ice/rock ratio in solar nebula analogs vary between 2 \citep{Lodders2003} and 4 \citep{Dodson-Robinson2009}, but may be increased by an order of magnitude or more near major icelines (see e.g. \citealt{Cuzzi2004}, and below). Second, the presence of icy or organic coatings on the surfaces of refractory grains can change the dominant collisional outcomes. Traditionally, the presence of a water ice mantle was expected to raise the sticking threshold velocity of aggregates by a factor of ${\approx}10$ \citep[e.g.][]{DominikTielens1997,Wada2013,Gundlach2015}, although more recently this canonical image of non-sticky refractory vs. sticky water ice grains has been called into question \citep{Kimura2015,Steinpilz2019}. 
Collision-induced phase-transitions leading to fusion/annealing can increase the stickiness even further \citep{Wettlaufer2010}. The expected jump in particle properties\footnote{A factor 10 increase in the sticking threshold translated to an increase in the maximum particle size that of a factor $10^2$ in turbulence-driven growth \citep{Birnstiel2012}.} across the water iceline then plays an important role in planetesimal formation models (see Sect.~\ref{sec:planetesimal_formation}) and may be used to locate the snowline indirectly using dust continuum emission \citep{Banzatti2015,cieza2016}. Organic coatings have also been suggested to increase particle stickiness at temperatures between 200 and 400 K \citep{Kouchi2002,Homma2019}, although recent experiments are not conclusive \citep{Bischoff2020}. Conversely, laboratory experiments have demonstrated that cold water ice ($T<200\mathrm{~K}$) behaves similar to bare silicates (\citealt{Musiolik2019}, and see \citealt{Kimura2020}). In addition, mixing in other, more volatiles ice species like CO$_2$ was also found the decrease the (effective) surface energy \citep{Musiolik2016}, suggesting the disk region where water ice enhances sticking is limited \citep[e.g.][]{Pinilla2017}.

Finally, even minor molecular species may play an important role through the process of sintering (essentially the fusing of grains at temperatures just below the sublimation point of one of their constituents). Operating near the snowlines of species with a range of abundances this process can turn aggregates more brittle and thus more susceptible to fragmentation \citep{Sirono2011,Sirono2017}, near the snowlines of species with even relatively minor abundances. Modeling by \citet{Okuzumi2016} has shown that the presence of sintered aggregates near snowlines in HL Tau can explain the radial structure observed in the dust continuum \citep{ALMA2015, Zhang2015}. More work is needed to understand the mechanical behavior of realistic, complex aggregates that are mixtures of different refractory and ice phases.

\subsubsection{Pebble dynamics and CHNOPS redistribution}\label{sec:pebble_chnops}
The formation and evolution of pebbles can have a profound effect on CHNOPS abundances and elemental ratios in a variety of disk reservoirs. 

First, the removal of small dust as a source of opacity from the outer disk and disk surface layers changes the disk temperature structure and radiation field, altering the freeze-out/condensation balance of (highly) volatile species \citep[e.g.,][]{Cleeves2016}. Second, efficient vertical settling can lead to sequestration of volatile species in the midplane, and thus their removal from the surface, through the "vertical cold finger" effect \citep{Meijerink2009,Krijt2016b,Xu2017,vanclepper2022}. Observational support for this comes from low abundances of C and O carrying species \citep{Kama2016,Du2017,Zhang2019,Zhang2020a}. In IM Lup, contrary to C and O, N does not appear to be depleted, suggesting a highly volatile carrier (N$_2$) that cannot readily be sequestered in this disk's midplane \citep{Cleeves2018}.

Arguably more important for planet formation is that pervasive radial drift will profoundly alter elemental and molecular abundances in the disk midplane, often concentrating material around snowlines \citep[e.g.][]{Cuzzi2004,ObergBergin2016,Booth2017,Stammler2017,Krijt2018,BoothIlee2019,Krijt2020}. Radially resolved CO depletion profiles inferred by \citet{Zhang2019} resemble model results when pebble settling and drift is included \citep{Krijt2018,Krijt2020}, and \citet{Zhang2020b} use this framework to estimate the pebble flux across the CO snowline as ${\approx}15{-}60~M_\oplus/\mathrm{Myr}$ in HD163296.

An unresolved question concerns the fate of ice-rich pebbles as they drift through major snowlines and begin to sublimate. For compact pebbles with a core/mantle structure, the drift timescale is shorter than the thermal desorption timescales for pebbles with sizes ${\gtrsim}1\mathrm{~cm}$, meaning ice-coated particles may drift considerable distances before fully sublimating, effectively dragging snowlines in by as much 40{-}60\% \citep{Piso2015}. For porous, more loosely bound aggregates, however, sublimation of major phases may lead to aggregate disintegration \citep{Aumatell2011}, possibly leading to an enhancement of small grains near e.g., the water snowlines \citep{Schoonenberg2017}. Recent laboratory experiments indicate aggregates may escape disruption if their ice content is below ${\approx}15\%$ or if they contain significant small dust grains \citep{Spadaccia2022}, although it is not clear how robust the surviving structures are against fragmentation in future collisions.

While the inner disk midplane is hidden from view, there are tantalizing signs of a link between the details of CNO chemistry in the inner disk and the behavior pebbles further out, with higher ratios of HCN/H$_2$O fluxes being found (on average) for systems with a more massive \citep{Najita2013,Najita2018} and/or more radially extended \citep{Banzatti2020} pebble disk. These results suggest that disks in which significant inward migration of pebbles has occured contain more water vapor (or, lower C/O) in their terrestrial planet formation regions. It is interesting to note that while IR line fluxes of e.g. HCN and C$_2$H$_2$ look very different around cooler M5-M9 stars \citep{Pascucci2009}, ALMA observations of CO, HCN, and C$_2$H indicate the (observable) chemistry around 10-100 au is similar in cool (M4-M5) stars compared to Sun-like stars -- although hydrocarbons may be more abundant around the cooler objects \citep{Pegues2021}. 

Despite the lack of a clear correlation between the locations of molecular rings and rings and gaps in seen in dust continuum \citep{jiang2022}, evidence is mounting that sub-structure in the gas and dust (Sect.~\ref{sec:disk_structures}) plays a role in shaping the observable chemistry in complex ways. For example, \citet{Facchini2021} present a resolved observations of e.g., CO, CS, HCN, and small hydrocarbons in the ringed and planet-hosting disk of PDS70, highlighting how emission of different molecules peaks at different locations relative to the dust continuum. Overall, the observed chemistry (e.g., the lack of SO) however points to a C/O$>1$. In contrast, \citet{Booth2021a} present detections of SO and SO$_2$ (but not CS) in Oph-IRS 48, suggesting C/O values below unity in this disk that contains an azimuthally asymmetric distribution of pebbles \citep{Booth2021a}. 
 For a larger sample of 26 protoplanetary disks, \citet{vanderMarel2021} find a tentative correlation between the C/O ratio (as traced by C$_2$H emission) and the location of the outermost pressure bump relative to the CO snowline location. 

Finally, the effects of radial drift may be mitigated by so-called `dust traps' in disks: radial locations where there are local pressure maxima that cause the drifting pebbles to pile up \citep{Pinilla2012a}. Dust traps are thought to cause the rings in millimeter-sized dust grains that form at the edges of disk gaps \citep{Pinilla2012b}, and they may also exhibit asymmetric pebble clustering \citep[e.g. Oph-IRS 48][]{vanderMarel2013a}. The presence of traps should produce a volatile depletion in the gas interior to the trap's location, in contrast with the volatile enhancement expected from radial drift. Such a trap-induced depletion in C-rich gas has been inferred from measurements in the inner disks of young Taurus stars, some of which are compact disks for which millimeter grain substructure has yet to be seen \citep{mcclur2019}. This ability of inner disk gas depletion to probe traps in known millimeter substructure was confirmed for TW Hya \citep{Bosman19}. If silicon gas is also measured, then the bulk solid composition of dust in the traps can be calculated and compared with Solar System bodies. For TW Hya, the dust has a C, N, O composition indicative of both CI chondrites and comets \citet{mcclure2020}.

\begin{deluxetable}{ c c c c c }

\tabletypesize{\scriptsize}


\tablecaption{Compilation of CHNOPS Carriers in the ISM and Protoplanetary Disks}

\tablewidth{0pt}


\tablehead{}
\tablehead{ \colhead{ Carrier } & \colhead{ $T_\mathrm{cond}\,$(K)} & \colhead{ $T_\mathrm{sub}\,$(K)} & \multicolumn{2}{c}{Approximate midplane snowline location$^{1}$ (au)}\\
& \emph{"Reset"} & \emph{"Inheritance"} &  \emph{Class 0 disk} & \emph{Class II disk}}

\startdata
Graphite &  & 2400 & ${\leq}0.05$ & ${\leq}0.05$ \\
Silicates & 1350 & 1350 & $0.11$ & $0.50$  \\
(Fe,Ni)$_{3}$P & 1000 & 1000 & $0.65$ & $0.15$  \\
FeS & 655 & 655 & $1.0$ & $0.23$  \\ 
CHON(S) &  & 400 & $1.8$ & $0.40$  \\
NH$_{4}^{+}$X$^{-}$ & & ${\approx}200{-}500$ & $3.9$ & $0.88$ \\
H$_2$O & $100{-}150$ & $100{-}150$ & $7.5$ & $1.7$ \\
NH$_3$ & 109 & 109 & $7.6$ & $1.7$  \\
CH$_3$OH & & 99 & $8.5$ & $1.9$ \\
PH$_{3}$ & 88 & 88 & $9.7$ & $2.2$ \\
H$_2$S & 54 & 54 & $17$ & $3.8$ \\
CO$_2$ &  & 53 & $17$ & $3.8$ \\
CH$_4$ & 26 & 26 & $38$ & $20$ \\ 

PO & 24 & 24 & $41$ & $18$ \\
CO & 23 & 23 & $43$ & $27$ \\
N$_2$ & 20 & 20 & $49$ & $36$ \\

\enddata
\tablenotetext{1}{\scriptsize{Locations based on the midplane temperature of an irradiated, viscously evolving \citet{Chambers2009} disk model around a young solar-mass star evaluated at $t=40~\mathrm{kyr}$ (Class 0) and 2 Myr (Class II).}}

\label{tab:CHNOPS_carriers}
\normalsize
\end{deluxetable}

\subsubsection{Planetesimal Formation: Where \& When}\label{sec:planetesimal_formation}

In modern planet formation theory, planetesimals form via a two-step process: pebbles are accumulated in dense clumps (or clouds) via the streaming instability (SI), and these clouds gravitationally collapse to directly form macrosopic planetesimals \citep[][Dr{\c{a}}{\.z}kowska et al., this volume]{Youdin2005, Johansen2014}. The initial mass function of planetesimals formed in this way peaks between 50 and a few 100 km \citep{Simon2016}, largely independent of pebble properties \citep{Simon2017}. This characteristic size of 100 km is supported by more recent theoretical considerations \citep{Klahr2020}, simulations \citep{Klahr2021}, and roughly matches observed kinks in the size distributions of asteroids \citep{Bottke2005,Morbidelli2009}. The gravitational collapse of a pebble clump is rapid but gentle \citep[e.g.,][]{Visser2021}, and it is customary to assume the primordial planetesimal essentially inherits the composition of the pebbles and ices present in that particular part of the disk. Observational support for planetesimal formation via pebble clump collapse comes from studying the orientation of Kuiper Belt binaries \citep{Nesvorny2019} and the structure of individual Kuiper Belt Objects like Arrokoth that have been visited upclose \citep{McKinnon2020}. 

Roughly since \emph{Protostars \& Planets VI}, perhaps inspired by \citet{Johansen2014}, hybrid models have become popular that connect disk-wide dust coagulation and drift to planetesimal formation via the streaming instability \citep[e.g.,][]{Krijt2016a,Drazkowska2016,Drazkowska2017,Carrera2017,Schoonenberg2017,Schoonenberg2018,Lenz2019}. In many such studies of the conditions needed to trigger SI, the required increase in dust-to-gas ratio (by a factor of ${\approx}2$, when starting from a solar metallicity; \citealt{Carrera2015}) was found to be too restrictive, and additional ways of "pre-concentrating" solids in specific disk locations (e.g., at a snowline) appeared to be necessary\footnote{\citet{Li2021} recently revised the metallicity threshold for SI down to below 0.01 (for selected pebble sizes), suggesting easier planetesimal formation across a range of disk locations, although the impact on disk-wide dust evolution \& planetesimal formation models has not yet been assessed.}. 

Depending on the method of pre-concentration, as well as specifics in the dust properties (e.g. sticking threshold and variation therein, porosity) and disk properties (initial surface density, turbulence), such hybrid models yield very different outcomes. For example, some models favour short bursts of planetesimal formation around the water snowline \citep{Schoonenberg2017}, others form planetesimals late and at large heliocentric distances \citep{Carrera2017}, and others still predict planetesimal formation over extended periods and essentially throughout the entire disk \citep{Lenz2019}. Recent studies also highlight the possibility of efficient planetesimal formation in existing pressure bumps \citep{Stammler2017,Carrera2021}. It is worth noting that many of these models rely on volatiles (usually H$_2$O) \emph{and} the radial drift of pebbles to locally increase dust-to-gas ratios via either: (1) retro-diffusion across the snowline followed by rapid freeze-out, and/or (2) a dramatic change the sticking behavior of grains near the snowline (see Sect.~\ref{sec:dust_pebble_planetesimal}). In disks with different temperature structures (for example those around M stars) these models would predict the main planetesimal formation zone to lie closer to the star \citep{Ormel2017}.

The locations where planetesimals and planets form can be probed implicitly through the radial positions of dust traps. Oph-IRS 48 is the best example of this, as the dust is both radially and azimuthally confined \citep{vanderMarel2013a}. However, for disks without resolved substructure, \citet{mcclure2020} have proposed that it is possible to locate planet-forming zones relative to different volatile snowlines using the relative depletion of different volatiles from inner disk gas. Specifically, any significant depletion of elements such as N, C, O, or Si suggests dust traps, but a low N/C or C/O abundance ratio implies traps beyond the \ce{N2} or \ce{CO2} snowlines, respectively. High O/Si abundance ratios are sensitive to traps outside versus inside the H$_2$O snowline. Future observations of inner disk gas with CRIRES+ and JWST will help to establish this technique.

An indirect way of timing widespread planetesimal formation in nearby protoplanetary disks is to compare their solid mass budget to the masses of mature planetary system \citep{Najita2014,Manara2018,Tychoniec2020,Mulders2021,Lovell2021}. While there are many uncertanties associated with converting continuum fluxes to dust masses \citep[e.g.][]{Zhu2019} these studies suggest planetesimal formation may begin early. Similarly, comparing spatial scales of observed pebble disks to planetary system demographics reveal dramatic inward mass transport must take place at some stage of planet formation, either in the form of early pebble migration, or later planet migration \citep{Mulders2021}. It is worth noting that while gas disks (as revealed by excess IR emission) live longer around cool stars \citep[][and Sect.~\ref{sec:disk_disperal}]{Carpenter2006, Luhman2012}, the total mass contained in pebbles (as estimated from submillimeter continuum emission) decreases faster \citep{Pascucci2016}, suggesting radial drift operates faster/differently \citep{Pinilla2020} and/or a faster conversion of pebbles into planetesimals.

\subsection{Planetesimal Evolution \& Protoplanet Accretion}\label{sec:protoplanet_accretion}

\subsubsection{Local planetesimal accretion}\label{sec:planetesimal_accreion}
Rapid runaway growth amongst small planetesimals results in the formation of bodies up to 100-km sizes on timescales of $10^{4-5}\mathrm{~yr}$ around 1 au \citep{Wetherill1989, KokuboIda1996, Weidenschilling1997}. There is a transition from runaway to oligarchic growth once the largest planetesimals dynamically excite smaller bodies in their feeding zones, which shuts down focusing \citep{Wetherill1989, KokuboIda1996, Weidenschilling1997}; this transition occurs roughly at a few 100 km in inner disk and ${\sim}10^3$ km in the outer parts \citep{Ormel2010}. Oligarchs continue to slowly but steadily gain mass by accreting smaller planetesimals \citep[][]{Kokubo1998,Rafikov2003}. After ${\sim}10^6$ yrs, the result is then a handful of Moon to Mars-sized planetary embryos, separated by 5-10 Hill radii and surrounded by a swarm of smaller planetesimals \citep{Kokubo1998}. Planetesimal accretion (in the absence of perturbing planets) can be thought as a fairly localized process, with feeding zones of growing bodies of the order of roughly 10 Hill radii \citep{Tanaka1999}, which, for an ${\sim}0.1M_\oplus$ at 1 au correspond to roughly 0.03 au. 

Moreover, planetesimal-planetesimal collisions during the runaway and oligarchic growth phases are not expected to lead to the loss of volatiles in hydrated minerals \citep{Daly2018SciA} or even ices \citep{Schwartz2018}, suggesting that compositional gradients in the planetesimal population (as shaped by formation and thermal evolution, see below) will be largely preserved during runaway and oligarchic growth. However, large numbers of successive embryo-planetesimal and embryo-embryo impacts may still lead to significant water loss, with simulations assuming perfect merging possibly overestimating the final water mass fraction by up to 10{-}20\% \citep{Haghighipour2022}.

\subsubsection{Planetesimal Thermal Evolution and $^{26}$Al}  
While planetesimal-planetesimal impacts contribute little to heating, the presence of short-lived radionuclides (SLR) can significantly alter the physical/chemical structure and CHNOPS budgets of planetesimals that form relatively early and/or large. In particular, $^{26}$Al's half-life of $\rm0.7\,Myr$ makes it a potent source of internal heating. 

Depending on how much a planetesimal's temperature rises, outgassing of highly-volatile ices (if present), followed by melting of water ice (if present), and finally partial and complete differentiation (i.e., the formation of a metal core and separation of lithophile and siderophile elements) can occur following melting of metal-sulfides and later silicates \citep[][]{Elkins-Tanton2012, Fu2014,Fu2017}. When heating is efficient (i.e., when planetesimals form early, are born larger, or the system just happens to have a high SLR abundance), silicate melting can lead to dramatic water loss at the planetesimal stage -- even if the planetesimal formation mechanism itself may have preferentially formed water-rich bodies \citep[e.g.][]{GrimmMcSween93,Lichtenberg2019,Lichtenberg2021a}. Similarly, volatiles like CO and CO$_2$ \citep{LichtenbergKrijt2021}, and more refractory carbon carriers can be lost, greatly reducing the bulk C/S ratio \citep{Hirschmann21}. 

The importance of formation age on parent body internal evolution is evidenced by the nature of the oldest meteorites in our collections: the iron meteorites, which were produced by metal-silicate segregation on a molten parent body following its accretion within just $\rm<300\,kyr$ (less than one half life of \ce{^{26}Al}) of CAI formation \citep{kruijer2014_science}. In contrast, the high porosities (i.e., low densities) of small (${\lesssim}100$ km) Kuiper Belt Objects suggests limited SLR heating, and hence planetesimal formation times of ${>}4$ Myr after CAI's in the outermost regions of the Solar System \citep{Bierson2019}.

\subsubsection{Pebble accretion and gas-driven migration}\label{sec:gas_migration}
As embryos continue to grow, the orderly picture painted above is complicated by orbital migration (the change in a planet's semi-major axis) and pebble accretion. 

The speed and even direction (i.e. inward vs. outward) of gas-driven Type I migration depends sensitively not just on planet mass \citep{Ward1997,Tanaka2002} but a slew of (local) disk properties including temperature and pressure profiles but also viscosity and details of heating/cooling/radiative transfer (see Paardekooper et al., this volume, for a review). In addition, convergent zones (where net torques on planets of certain masses vanish) can act as planet traps \citep[e.g.,][]{masset2006,Lyra2010,Dittkrist2014}. Heat transitions, major snowlines, and dead zones can all act as planet traps \citep{Hasegawa2011, Bitsch2015a}, and their locations are expected to move as disks evolve. In addition, excitation of orbital eccentricities from planet-planet interactions affects the planets' torque balance and drastically reduces the effectiveness of convergence zones \citep{cossou2013,izidoro2017}. While outward migration may dominate for low-mass cores in disks with low viscosity \citep{Speedie2022}, inward migration is still expected for for planets with masses ${\lesssim}M_\oplus$ around Sun-like stars \citep{Bitsch2015b}.

Pebble accretion plays a major role in modern theories of planet formation \citep{Ormel2010,Lambrechts2014}. For our Solar System, there appears to be consensus on pebble accretion playing a role in the rapid growth of the giant planets \citep{Lambrechts2012, Levison2015, Johansen2017, Alibert2018}. The formation of the terrestrial planets does not appear to require pebble accretion per se \citep[see also][]{Chambers2016,Nimmo2018,Izidoro2021}, perhaps because pebble accretion was less efficient inside the water snowline due to the smaller pebble sizes \citep{Morbidelli2015}, or because Jupiter blocked the influx of pebbles altogether as early as ${\approx}1\mathrm{~Myr}$ (Sect.~\ref{subsec:accreting_earth}). Nonetheless, some recent studies argue pebbles played an important role in the formation of Venus, Earth and Mars \citep{Johansen21}.

For extra-solar systems, many studies have focused on modeling the formation of Super-Earths and Mini-Neptunes, highlighting the interplay between pebble accretion and planet migration \citep[including][]{Bitsch2019,Coleman2019,Schoonenberg2019,Izidoro2021chains}. Within this framework, emerging planetary architectures in the terrestrial planet region depend sensitively on the (evolution of the) pebble mass flux \citep{Lambrechts2019,Drazkowska2021}. The formation of systems like TRAPPIST-1 around low-mass stars may be explained via a combination of planetesimal formation near the water iceline, efficient pebble accretion until the pebble isolation mass is reached, followed by Type I migration and the formation of resonant chains \citep{Ormel2017,Coleman2019,Schoonenberg2019}.

\subsubsection{Role of Giant Planets \& Disks Substructure}\label{sec:disk_structures}
The sizes of pebbles are roughly similar to the operating wavelengths of interferometers like ALMA, which means their spatial distributions in nearby protoplanetary disks can be studied down to spatial scales of ${\lesssim}5-10$ au, revealing the pervasive and dramatic presence of so-called substructure (e.g., \citealt{Andrews2020}; Bae et al., Manara et al., this volume), even in very young protoplanetary disks \citep{ALMA2015,segura-cox2018}. Although other origins are not easily ruled out, the deeper gaps in the radial distribution of pebbles are usually attributed to the presence of forming giant planets \citep{Huang2018,Zhang2018}. The direct detection of PDS70b offers convincing evidence of this happening in at least one system \citep{Keppler2018, Benisty2021}.

Given the importance of the pebble flux on the growth of the inner planets \citep{Lambrechts2019}, it is reasonable to assume the (early) formation of giant planets on wide orbits has a profound influence on slowing down the growth of the inner planets by starving them of pebbles. And indeed, the stellar mass dependent occurrence rates of giant planets, super Earths, and extended (i.e. structured) vs. compact disks seem to support such a picture \citep{vanderMarelMulders2021}. \citet{Speedie2022} propose that this observed split between compact and large, structured disks is caused by diverging giant planet migration pathways in high vs. low viscosity disks. The efficiency with which giant planets can block the flow of solids depends sensitively on planets mass, disk viscosity, and the dust/pebble size distribution \citep[e.g.,][]{Bae2019}. In the fragmentation-limited region of the disk (Sect.~\ref{sec:dust_pebble_planetesimal}), solids will frequently change size and thus aerodynamic behavior \citep[e.g.,][]{Misener2019}, suggesting fragments can still carry mass and solids CHNOPS through the gap.

There are several other mechanisms that become important in the presence of giant planets. First, by blocking the inward flux of solids and therefore ices, the appearance of giant planets on the scene effectively freezes in the current snowline location, creating what is known as a \emph{fossilized} snowline \citep{Morbidelli2016}. The idea here is that the barrier thrown up by a giant planet (outside the snowline) results in the gas arriving in the inner disk through advection being depleted in water vapor. 

Second, giant planets themselves are not static but migrate inward via Type II migration at a velocity roughly equal to the viscous speed of the disk \citep{Lin1986}. Thus, the planet population inferred from e.g. the DSHARP sample may look quite different once these planetery systems mature \citep{Lodato2019}. If giants planets form early, and the disk sticks around for long enough (see next Section), one or more giant planets can migrate through the terrestrial planet-forming region, which can disrupt the growth of rocky planets and drastically widen their feeding zones \citep{Raymond2006,Mandell2007}.

In this context, it is interesting to highlight uncertainties in the early growth of Jupiter's core; with some studies favouring the water snowline as the formation region \citep{Morbidelli2012}, and others favouring core formation outside \ce{N2} iceline followed by inward migration to explain the enhancements of C, N, S, P, Ar, Kr and Xe in Jupiter's atmosphere \citep{ObergWordsworth2019,Bosman19}. Irrespective of its role in separating reservoirs of material, the need to form Jupiter before the dissipation of the gas disk strongly supports the role of pebble accretion in the outer solar system, as approaches relying on accretion by planetesimals alone largely failed to work rapidly enough \citep[e.g.,][]{Levison2010,Levison2015,Johansen2017,Alibert2018}.

\subsubsection{Gas Disk Dispersal \& Primordial Atmospheres}\label{sec:disk_disperal}
Eventually the outer disk gas either accretes or dissipates (Pascucci et al. this volume). The inevitable exit of the primordial gas disk does not signal the end of terrestrial planet formation (next Section) but does bring a halt to important processes like pebble accretion and Type I and Type II migration. Average dispersal timescales are around $2{-}4$ Myr \citep{Lada2006, Mamajek2009}, with gas disks around low-mass stars sticking around longer \citep{Carpenter2006,Luhman2012}, while disks around stars with ${>}2M_\odot$ evolve faster \citep{Ribas2015}. The short lifetimes however may be attributed to a bias towards dense star forming environments, with \citet{Pfalzner2014} finding higher disk fractions in co-moving groups, and arguing that  30\% of all field stars can host disks for more than 10 Myr. Recently, \citet{Michel2021} argued lifetimes of closer to 8 Myr may be more representative for disks with considerable substructure. In this context then the Solar nebula, with a lifetime of ${\approx}4$ Myr (\citealt{Wang2017} and Sect. 2.2) does not appear to be unusual, although a considerable spread in ages -- possibly influenced by the evolution of substructure and environment -- is found.

If planetary embryos manage to grow large enough before the disk dissipates, they may be able to hold on to a (primordial) hydrogen and helium dominated atmosphere. These atmospheres can play a role in shaping CHNOPS budgets by processing any incoming pebbles \citep{Johansen21}. \citet{LeeChiang2015} estimate that embryos with masses ${\sim}0.5{-}0.8 M_\oplus$ are capable of binding atmospheres of ${\sim}10^{-4} {-} 10^{-2}M_\oplus$ (see also Sect.~\ref{subsec:accreting_earth}). In systems where icy pebbles enriched the inner disk gas in volatiles, these primordial atmospheres may have had molecular and elemental abundances (and abundance ratios) very unlike the primordial nebula (\citealt{BoothIlee2019,Banzatti2020} and Sect.~\ref{sec:pebble_chnops}). For example, \citet{Bitsch2021} show how wet mini-Neptunes can form wholly inside the water snowline just from the accretion of inner disk gas enriched in water.

\subsection{Endgame: Final assembly of terrestrial planets}\label{sec:endgame}

\subsubsection{Dynamical (In)stability}\label{sec:dynamical_instability}

It has been argued that the two key dynamical processes shaping the orbital architectures of planetary system are migration and dynamical instability. Starting from a universal set of planetary precursors, just those two ingredients -- migration and instability -- can broadly explain the diversity of planetary systems, from the Solar System to giant exoplanets and systems of close-in `super-Earths' \citep[see extensive discussion in][]{Raymond2020}.

Analysis of systems containing multiple close-in low-mass planets discovered by the {\it Kepler} mission has found that such systems on average tend to be spaced, in units of Hill radii, more widely than expected \citep{FangMargot2012,PuWu2015ApJ}.  This is interpreted as the signature of a late phase of giant impacts, as decades of simulations of late-stage accretion of the terrestrial planets have shown that spacing by Hill radii is the characteristic outcome of a late collisional phase \citep[see reviews by ][]{Morbidelli2012,Raymond2014}. Systems of giant exoplanets also show clear signs of dynamical instability.  Their broad eccentricity distribution is naturally matched if 75-95\% of all giant exoplanet systems are the survivors of planet-planet scattering, triggered by dynamical instability in the high escape-speed regime \citep{Ford2008,Juric2008,Chatterjee2008}.  This model is well-accepted as it is extremely robust, deceptively simple, and able to explain several other aspects of the dynamical configurations of giant exoplanet systems \citep{Raymond2010}. Likewise, to match the period ratio distribution of close-in small planets, 95\% or more must be the survivors of a late phase of collisions \citep{izidoro2017,Izidoro2021chains}.  Such instabilities among giant planets, at least within a few au, would have a devastating effect on the growth of rocky planets in the same systems \citep{Veras2006,Raymond2011,Raymond2012} and in some cases close-in super-Earths \citep{mustill2017}.

There is a plausible narrative that can explain why late instability are likely common, built on the processes of orbital migration and dynamical instability \citep[for more detail see][]{Raymond2020}.  The narrative goes as follows.  Planetary embryos and giant planet cores cannot avoid orbital migration as long as they grow sufficiently massive early enough in the gaseous disk phase \citep{Kley2012,Baruteau2014}. Many generations of hydrodynamical simulations of migrating protoplanets show that they are naturally organized into chains of mean motion resonances, in which neighboring planets' orbital periods form the ratio of small integers such as 2:1, 3:2, 4:3 and so on \citep{Cresswell2007,Pierens2013}. Many resonant chains likely migrate all the way to the inner edge of the gaseous disk, where a strong positive torque acts as a migration barrier \citep{masset2006,romanova2006}. This is thought to be the case for low-mass planets such as the progenitors of `super-Earths' and `sub-Neptunes' \citep{terquem2007migration,Ogihara2009,cossou2014}.  Other migration halting mechanisms may exist to trap resonant chains of planets at larger orbital radii; for instance, the HR8799 system consists of four mega-Jupiters in a chain of resonances extending from ${\sim}18$~au to ${\sim}70$~au \citep{Marois2008,Marois2010,Fabrycky2010}.

\subsubsection{Giant impacts \& Late Accretion}\label{sec:giant_impacts}

The late phases of the growth of terrestrial planets are thought to have been characterized by a series of giant impacts between roughly Mars-mass or larger planetary embryos \citep{Morbidelli2012,Raymond2014}. These giant impacts had the potential to alter the planets' bulk compositions by preferentially stripping the outer layers, leading to erosion of a growing planet's atmosphere, oceans, or even mantle \citep{Genda2005,Asphaug2006}. Yet there is a broad spectrum of outcomes of such impacts \citep{Genda2012,Leinhardt2012}, and certain configurations may lead to the formation of a large satellite such as the Moon \citep{Benz1989,Canup2001}. The stochastic nature of this phase effectively erases much of the previous phases of planetary growth such that the nature of the Moon-forming impact cannot be used to trace back our system's earlier history.

Late planetesimal impacts can have a strong erosive effect on a planet's atmosphere and water budget \citep{Schlichting2015,Schlichting2018}, with habitable zone planets around M stars being more susceptible to impact stripping \citep{Wyatt2020}. While the Earth's late veneer likely had a local origin (Sect.~\ref{subsec:accreting_earth}), a substantial amount of volatile rich material from the outer system may be directed towards terrestrial planet formation regions in systems with different planetary architectures \citep{Marino2018}. Finally, terrestrial planets in relatively old gas-rich debris disk systems may be able to accrete some of the secondary C and O rich gas directly \citep{Kral2020}. As discussed earlier, P was likely delivered to rocky planets in the solar system as schreibersite, (Fe,Ni)$_{3}$P. The P from schreibersite can be made available for life by hydration or by lightning-induced chemistry \citep{Hessetal2021}.

\begin{figure*}[t!]
\centering
\includegraphics[clip=,width=.95\textwidth]{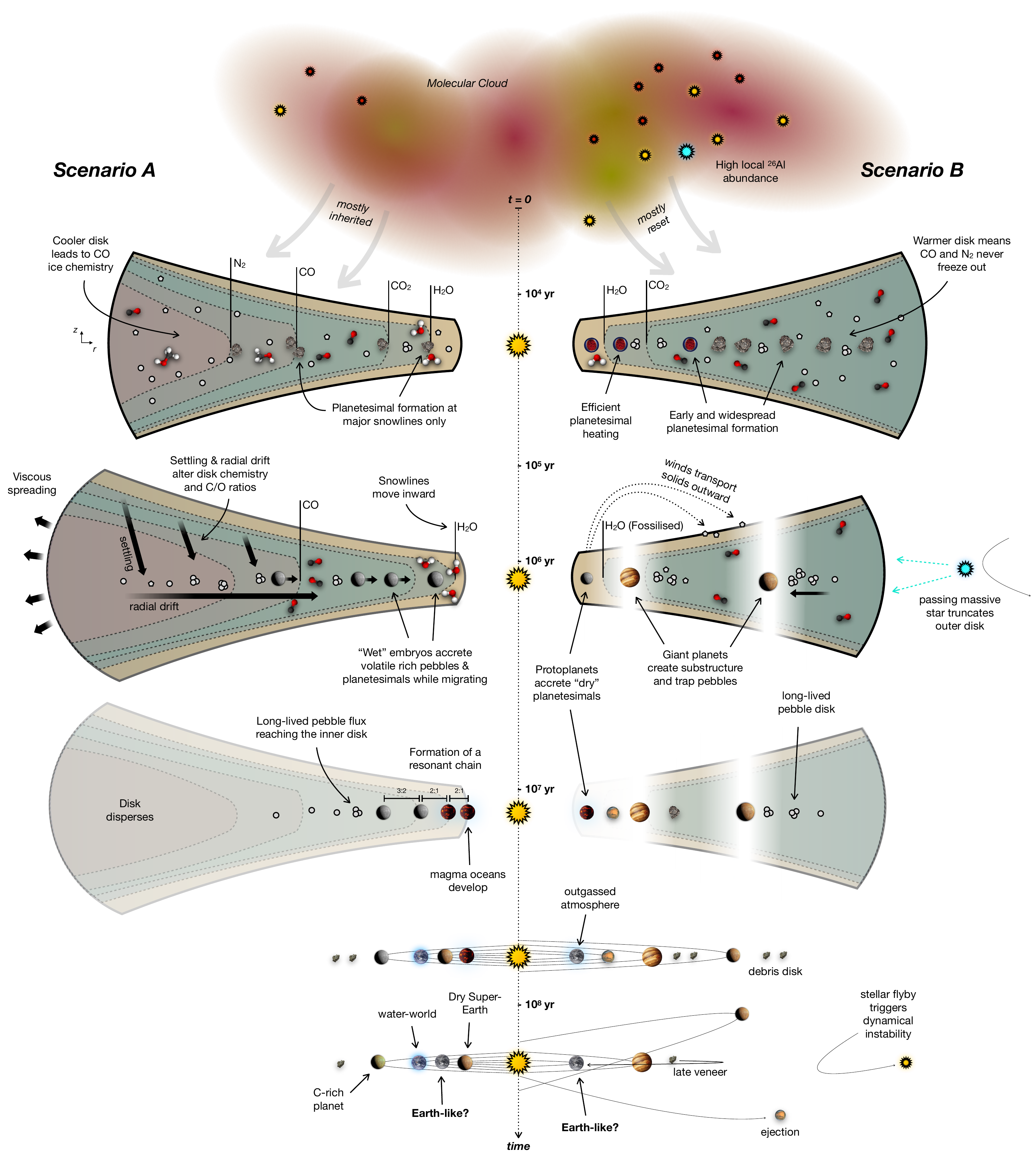}
\caption{Illustration of two hypothetical scenarios leading to the formation of diverse planetary systems and planets with varying degrees of chemical habitability. Individual processes and key branching points are discussed in more detail in Sects. \ref{sec:planet_formation} and \ref{sec:branching_points}. In Scenario A, a cold disk forms in a relatively quiet part of a star forming region, and planetesimal formation occurs exclusively at major snowline locations. Planetary embryos accrete volatile-rich planetesimals (this particular system formed without much $^{26}$Al) and pebbles, while themselves migrating inward to form a resonant chain. This system does not become dynamically unstable, but variations in distance to the star and ongoing geophysical processes (e.g., the duration of the magma ocean phase, and ability of the core to sequester volatiles) leads to variations in the types of planets that form. In Scenario B, the disk forms with a higher metallicity and SLR abundance, and planetesimals are able to form throughout. The formation of giant planets causes substructure, fossilizes the water snowline, and shuts off the radial pebble flux early on. Terrestrial planets form from dry, de-volatilized planetesimals, but may still acquire CHNOPS later on during externally triggered instability and late veneer-like event.}\label{fig:BP_cartoon}
\end{figure*}

\section{BRANCHING POINTS FOR MAKING CHEMICALLY HABITABLE WORLDS}\label{sec:branching_points}

We now speculate about what we consider to be major branching points in the formation of chemically habitable planets, using Fig.~\ref{fig:BP_cartoon} to illustrate diverging pathways.

\subsection{A note on stellar environment}
While planet formation is often discussed in the context of isolated protoplanetary disks, we recognize that the stellar \emph{environment} can play an important if indirect role at several key stages during the planet formation process. \citet{Parker2020} reviews three key stellar density thresholds: First, above ${\sim}10~M_\odot~\mathrm{pc^{-3}}$, FUV and EUV radiation from nearby massive stars can affect protoplanetary disks. At stellar densities ${>}100~M_\odot~\mathrm{pc^{-3}}$ planetary orbits (i.e., eccentricitiy, inclination, and semi-major axis) can be altered, and in the densest regions with  ${>}1000~M_\odot~\mathrm{pc^{-3}}$ protoplanetary disks themselves can be truncated by close stellar encounters. Proximity to nearby high mass stars increases the density threshold required to form ices in dense molecular clouds, with less strong ice absorption towards Ophiuchus, which is exposed to the irradiation from A-stars in upper Scorpious, than towards the less-exposed Taurus molecular cloud \citep[see discussion on p.89 of][]{mcclure2010, williams1992}. Moreover, proximity to short-lived massive stars can play a major role in setting the SLR abundance, greatly affecting volatile losses during the planetesimal stage. 

Within the vicinity of the Sun (500 pc), the Orion Nebula Cluster is likely the only star forming region that may have met or exceeded the third density threshold.  \citet{Winter2020b}, by identifying old co-moving groups in Gaia data, find planetary system architecture (in this case the median semi-major axis and orbital period) differ in stellar clusters versus field stars, with, e.g., Hot Jupiters existing primarily in stellar phase space overdensities. (However, see also the independent analysis of the Winter et al. data by \cite{Adibekyan2021}, which comes to different conclusions when considering a smaller but less biased sample.). And yet, planets are predicted to survive \citep{Fujii2019} and have been observed in open clusters \citep[e.g.,][]{Quinn2012,Meibom2013, Livingston2018,Curtis2018}.

\subsection{Stellar metallicity and CHNOPS abundances}
At the most fundamental level, the metallicity and relative elemental abundances protostellar system control the amount of solids and CHNOPS available for planet formation.

Metallicity clearly plays an important role in planet formation. On the one hand, an environment with a metallicity that is too low may prohibit any planet formation altogether. Protoplanetary disks with low metallicities have shorter lifetimes (due to increased efficiency photoevaporation) \citep{Kornet2005, Yasui2010,Ercolano2010}, decreasing the chances for planet formation to occur, and theoretical models of planet formation in circumstellar disks predict that Earth-mass planets would only start to appear around host star metallicities of -1.0 dex \citep{JohnsonLi2012, Mordasini2012}. 

On the other hand, an environment with a metallicity that is high is more likely to form giant planets. These have been suggested as detrimental to small planet formation (or at least chemically habitable planet formation) due to the influence of the giant planet's eccentricity and/or migration on the terrestrial planet's long-term dynamical stability \citep[e.g.,][]{Dressing2010, Kita2010}, or even prohibitive of the formation of a terrestrial planet in the first place \citep[e.g.,][]{Raymond2011,Lambrechts2014, Izidoro2015, Owen2018}. \citet{Owen2018} observe that super-Earths ($\sim$terrestrial planets) with long periods are more common around lower metallicity stars (which host fewer giant planets), consistent with fewer solids resulting in cores not massive enough to accrete large H/He envelopes and/or formation after the disk dispersed. And \citet{BoothOwen2020} flipped the script on the ``solar twin''--small planet connection first proposed by \citet{Melendez2009}, instead suggesting that the 10\% refractory element depletion seen in $\sim$10-20\% of solar twin stars is due to trapping of dust exterior to the orbit of a forming giant planet, preventing refractory-rich material from accreting onto the star. However, there is evidence from both the \textit{Kepler} sample and the combination of direct imaging and long-term radial velocity (RV) monitoring that longer period giant planets are \textit{highly likely} to be accompanied by an inner small (${\lesssim}4R_{\oplus}$) planet \citep{Zhu2018, Bryan2019}, although this evidence is complicated by the recent survey of giant planet hosting-systems by \citet{Barbato2018} that found no inner super-Earth companions.

Thus far, however, few obvious differences have been observed in the detailed compositions stars known to host (mostly large) planets versus those not known to host planets, beyond that first observed in [Fe/H] and an enhancement in alpha elements when [Fe/H] is low (Sect.~\ref{sec:exoplanet_hosts}). With the advent of both space-based all-sky surveys for planets (e.g., TESS, PLATO) and ground-based all-sky stellar spectroscopic surveys (e.g., APOGEE, LAMOST, GALAH, WEAVE, 4MOST, MOONS), larger-scale and/or more precise comparisons between the compositions of stars hosting different types of planets, particularly potentially-terrestrial ones, will be of great interest to the context of chemical habitability. If the the interior or surface compositions of rocky planets depend on their host star abundances, this may also be observationally constrained via their atmospheric abundances, although this is a relatively new area of research \citep{Herbort2020,Bitsch&Battistini2020}.

\subsection{Planetesimal Formation \& Evolution}

Despite the success of the streaming instability model of planetesimal formation to overcome the meter-size barrier, uncertainties in the timing and location of planetesimal creation constitute a major obstacle in our understanding of how, where, and what types of planets emerge inside protoplanetary disks. Snowlines and their evolution in time play an important role by (1) influencing the volatile budget of dust and pebbles and (2) being prime sites of efficient planetesimal formation. Indeed, it is possible that, rather than a continuous disk, two or three spatially-separated rings of planetesimals formed, associated with the condensation fronts of silicates, water and CO \citep{Morbidelli2022,Izidoro2022}. If planetesimal formation indeed follows snowline locations, we may expect similar intra-system compositional variability to emerge in different systems\footnote{Interstellar interloper 2I/Borisov was observed to be very CO rich \citep{Cordiner2020,Bodewits2020}, prompting comparisons to the unusual comet C/2016 R2 \citep{Biver2018} and theories about an origin near the CO iceline \citep{Mousis2021,Price2021}.}. While highlighting the role of snowlines, it is interesting to note that their locations (during the protoplanetary disk phase) relative to the location of the classical habitable zone (for mature systems) are strongly stellar mass dependent. For low-mass stars, for example, the habitable zone is located considerably further inside the water snowline \citep[e.g.,][]{Mulders2015b,Desch2020}, suggesting the building blocks forming near what would become the habitable zone were originally much drier.

Another major source of inter-system variability can come from variation in SLR abundances\footnote{There could also be more spatial variation in the $^{26}$Al distribution in the same disk: Recent work by \citet{Adams2021} suggests that it may be common for planet-forming material close to the the magnetic truncation radius of star/disk systems ($\sim$0.1 AU) to have $^{26}$Al abundances enhanced over early Solar System values by factors of $\sim$10-20 due to spallation reactions from stellar cosmic rays.}. In systems with considerable amounts of SLR, heating followed by the loss of volatiles from planetesimals that form early and/or are born big can be very dramatic. In these systems we expect planetesimals to be water, C, and N poor, while losses of P and S may be less severe as they exhibit more siderophile/chalcophile behavior during planetesimal differentiation.

There exist few direct constraints on planetesimal CHNOPS compositions in extra-solar systems. Possible avenues are studying second-generation gas originating in planetesimal belts in debris disks \citep[e.g.,][]{Marino2016, Matra2017,Wyatt2020}, or constraining the composition of material raining down on polluted white dwarfs \citep[e.g.,][]{Jura2014,Xu2014,Xu2017,Doyle2019}. So far, these constraints paint a picture of planetesimal compositions reminiscent of the asteroid and Kuiper Belt populations in our solar system. In the near future, JWST, especially in combination with ALMA, will be instrumental in further constraining inner disk chemistry and elucidating its link to the large-scale redistribution of pebbles. Further detailed spectroscopic investigations of the $\sim$500,000 new white dwarf candidates from Gaia DR2 \citep{GentileFusillo2019,Melis2018, Kaiser2021} will likely greatly increase the sample of polluted white dwarfs and thus constraints on the diversity of extra-solar planetesimal CHNOPS budgets.

\subsection{Pebble vs. Planetesimal Accretion}

In terms of emerging planetary CHNOPS budgets, planetesimal and pebble accretion pathways involve very different feeding zones and loss mechanisms for CHNOPS elements: Planetesimals are mostly accreted locally and can devolatilize following internal heating and differentiation. Following the dissipation of the gas-disk, mixing and transport of material is still possible, albeit by less efficient dynamical mechanisms (Sect.~\ref{sec:endgame}). Conversely, while pebble accretion is quite fast it is not particularly efficient and would require potentially hundreds of Earth masses of pebbles to build for example the Solar System's terrestrial planets. Such mass is not expected to be available exclusively within the inner regions of protoplanetary disks, and therefore a large influx of material from more distant regions of the disk would be required, potentially flooding the terrestrial planet formation region with water and other volatiles.

The radial pebble flux reaching inner protoplanetary disks can be decreased or completely shut down by changing pressure gradients in the disk \citep{haghighipourboss2003}, where possible mechanisms to stop the flow from the outer disk to inner disk include the growth a giant planet core \citep{Kruijer17}, or pressure bumps preceding giant planet growth \citep{Brasser2020}, so it is possible to have pebble accretion in one place (or at a certain time), but a lack of pebbles elsewhere. 

Simulations of rocky planet formation accounting for planetesimal accretion offer different predictions about the role giant planets in terrestrial planet composition, for example: the scattering of planetesimals following Type II Jovian planet migration results in terrestrial planets with high water fractions \citep{Raymond2006b,CarterBond2012}; higher eccentricities of Jovian planets reduce water delivery to terrestrial planets \citep{Raymond2009}; and inner super-Earths with outer giant companions will be volatile-poor and have higher densities \citep{Schlecker2021}.

There has been less work investigating emerging rocky planet compositions in pebble accretion-dominated scenarios. Even though pebbles will originate from the outer disk regions, and their journey inward will have them traverse multiple snowlines (e.g. Table~\ref{tab:CHNOPS_carriers}), the standard assumption is that their relatively small size will not allow them to hold on to most volatile species for very long \citep{Piso2015}. Thus, while they may alter the gas-phase abundances in the inner disk (see Sect.~\ref{sec:pebble_chnops}), the pebbles themselves will be fairly "dry" once they reach the inner disk. Nonetheless, in the absence of a barrier in the outer disk, the terrestrial planet formation region may be flooded with material rich in water ice \citep{Bitsch2021} and refractory/organic carbon, erasing any (prior) carbon loss that occurred via oxidation or photolysis \citep{Klarmann2018}. Still, the H$_2$/He envelopes of early Earth and Mars analogues (see also Sect.~\ref{sec:disk_disperal}) may be hot and dense enough for organic molecules to undergo pyrolysis and sublimation during pebble accretion, allowing the carbon and oxygen to diffuse back to the protoplanetary disk and resulting in relatively dry and carbon-poor planets \citep{Johansen21}.

With the increase in exoplanet detection and characterization capabilities, these predictions are now becoming testable via detailed, careful comparisons of small planets with and without known giant planet companions.

\subsection{Migration \& Instabilities}
Two major lessons learned in the last few decades are that planets do not necessarily form where we find them today and that potentially violent dynamic instabilities are likely the norm rather than exception (Sect.~\ref{sec:dynamical_instability}). Even with perfect knowledge of a system's initial conditions (which elements primarily in what form in which regions of the disk) and detailed knowledge of the primary modes of accretion, planets can thus move great distances and then combine with each other in chaotic fashion in a way that can frustrate reverse engineering the provenance of each. Close-in super-Earths are a good example of this phenomenon, as models for their starting location span from the innermost parts of the disk \citep{ChatterjeeTan2014} to far beyond the snow line \citep{terquem2007migration,Izidoro2021chains}. 

A number of lines of evidence -- from the giant planets' orbits and small body populations -- suggest that our Solar System also underwent a dynamical instability \citep{Nesvorny2018ARAA}. Our current understanding of the instability indicates that the giant planets likely scattered off of each other, perhaps even ejecting a primordial additional ice giant.  However, the Solar System's instability was gentle in comparison with that experienced by most giant exoplanet systems, as Jupiter and Saturn never underwent a close encounter; if they had, our Earth would not have survived \citep[see][]{Raymond2011}.

Planetary migration during the gas-rich protoplanetary disk phase, especially when crossing (major) snowline locations, can lead to planetary compositions that may be unexpected purely from the planet's final location. For example, if planetary cores form near or outside the water ice-line, planets ending up well within the snowline location may hold large reservoirs of water, with water mass fractions of up to ${\sim}10$ wt\% \citep[e.g.][]{Raymond2008,Ogihara2009,Bitsch2019}, although migration can also create a diversity of super-Earth compositions including purely rocky planets \citep{Raymond2018,Izidoro2021chains}. \cite{Bitsch&Johansen2016} showed that decreasing the water-to-silicate ratio of dust grains in the disk from 1:1 (as in the pebble accretion model of \citealt{Bitsch2015b}) to 1:3 resulted in more migration of icy cores to sub-AU orbits, as lower water content decreases the temperature gradient near the ice line thereby reducing the chances of outward migration; this implies that some short period super-Earths, perhaps those around oxygen-poor stars, should contain larger fraction of water accreted from beyond the ice line (see also \citealt{Bitsch2021}). For compact systems around M dwarfs, the formation scenario outlined above predicts very wet planets with water contents ${\sim}10\%$ \citep{Unterborn2018,Schoonenberg2019}. For TRAPPIST-1, density constraints for the outer 4 planets (e{-}h) allow up to ${\approx}5\%$ water if they sport an Earth-like core and mantle, but other solutions that allow higher water contents cannot be ruled out \citep{Agol2021}.

While the radius-period distribution of short period super-Earth/mini-Neptune planets is consistent with planet formation models that result in predominantly ``rocky'' versus icy compositions \citep[e.g.,][]{OwenWu2013,OwenWu2017,Jin&Mordasini2018,Rogers&Owen2021}, in some contexts rocky planets may have up to 20\% water content by mass \citep{Gupta&Schlichting2019}. The true diversity of super-Earth/mini-Neptune compositions is an important issue that will be informed by additional precise mass measurements, atmospheric composition constraints, and better constraints on the occurrence rate of small planets beyond $\sim$30 days \citep{Lee2022}.

Resonant chains are generally stable during the gaseous disk phase.  In some cases they may break -- for instance, leading to a collision between neighboring low-mass planets close to their star -- but if the gas disk is still present the resonant chains can simply re-form.  After the dissipation of the gaseous disk, a large fraction of resonant chains become dynamically unstable.  The instability trigger may be chaotic overlap of orbital resonances \citep{Batygin2013}, interactions with a remnant disk of planetesimals \citep{Levison2011}, or even the dispersal mechanism of the disk itself \citep{liu2022}. There are hints that external triggers for instability may play a role, for instance from stellar flybys \citep{malmberg2011} or from long-term perturbations from wide binary stars \citep{Kaib2013Nature}. An interesting opportunity is then presented by systems in which resonant chains of multiple planets have been able to survive. For example, \citet{Raymond2022} used the current resonant architecture of the TRAPPIST-1 system to show that each planet accreted at most $10^{-4}{-}10^{-2}M_\oplus$ after the disk dissipated, implying that any large reservoirs of e.g., water must have been incorporated already during the early formation (see above).

\subsection{CHNOPS in the Core, Mantle, and Atmosphere}
Moving beyond bulk abundances, the partitioning of CHNOPS elements between the core, mantle, surface and atmosphere is controlled by geophysical processes associated with the later stages of planet formation. The main avenues for removing CHNOPS from the (near) surface are loss to space (through outgassing or through impacts) and sequestration in to the core. Replenishment can occur via planetesimal impacts after core formation and magma ocean solidification (see below). The role these processes played in shaping the Earth is discussed in Sect.~\ref{sec:Earthbackintime}.

Large amounts of volatile elements can be sequestered in the metallic core during core formation (e.g. Tables \ref{tab:chnops_earth1} \& \ref{tab:chnops_earth2} and Fig.~\ref{f:evo_dist}), potentially leaving the mantle highly depleted in siderophile and chalcophile elements while lithophile species largely stay behind. The process of core formation therefore plays an important role in shaping the mantle's composition and oxidation state, influencing for example the composition of secondary outgassed atmospheres. Generally, more reducing conditions increase the amount of H and C entering the core, while N and S become increasingly lithophile \citep[e.g.][]{grewal2019_gca,Desch2020}. The detailed partitioning of elements during core formation depends sensitively on pressure and temperature, but also on composition -- e.g. oxygen fugacity.

Planetary-scale magma oceans may develop already during the protoplanetary disk phase, or following giant impacts like the Moon-forming event, and can take between ${\sim}1{-}100$ Myr to fully crystallize depending on its composition and distance from the star \citep[e.g.,][]{Hamano2013}. The specifics of the crystallisation process the nature of the primordial/secondary atmosphere will determine the partitioning of available CHNOPS elements between the atmosphere, upper/lower mantle and core. If still embedded in the primordial protoplanetary disk, significant amounts of hydrogen may be ingassed, while atmospheric escape can become important later on\footnote{For planets orbiting M stars in particular, water loss from the surface during the magma ocean or runaway greenhouse phase on HZ-like orbits around M dwarfs may be significant \citep[e.g.,][]{Luger2015}.}. These processes are punctuated by small and large impacts (see previous sub-section), complicating things further. Solubilities of CHNOPS elements depend on the composition and oxygen fugacity of the melt. Generally, C solubility is lower than water's, and N is less soluble than both of those. S instead is very soluble in silicate melts, although the behavior is a complex function of temperature, pressure, etc. For reduced mantles, C and N (and S) stay in the mantle. For oxidized mantles, C outgasses as CO$_2$ (before water does), N outgasses as N$_2$, and S stays in the mantle \citep[e.g.,][]{grewal2019_gca,Unterborn2020,Desch2020}. A detailed discussion of the geophysical evolution of planetary-scale bodies is presented in Lichtenberg et al. (this volume).

The variation in outcomes during these potentially crucial stages of terrestrial planet construction is difficult to predict accurately because of uncertainties in the behavior of CHNOPS carriers across large variations of temperatures, pressures, redox states, etc. However, the outcome of such processes is influenced greatly by the CHNOPS budgets as set during early stages, e.g., those leading up to and following planetesimal formation.

\subsection{Consequences of Compositional Variability}
CHNOPS are essential for life and many of the climatic and geodynamic characteristics of the modern Earth (Sect.~\ref{sec:Earthbackintime}). Yet, these elements are often present at trace quantities in the environment and early in a planet's life have many possibilities for loss to space or terminal sequestration into its interior (Fig.~\ref{f:evo_dist}). Given this, it is tempting to focus on how fortunate Earth is to have acquired enough of these vivifying elements. Whilst this view is important, it is also the case that larger quantities of CHNOPS could cause severe problems for the chemical habitability of a planet, and the origin of life. Given that alternative routes of planet formation can lead to significantly increased CHNOPS contents, we here briefly comment on the outcome for chemical habitability when CHNOPS are much more abundant than on Earth.

\textit{Carbon and oxygen: }Carbon alone has a clear role in setting planetary climate.  However, silicate weathering offers the potential to remove vast quantities of carbon present as \ce{CO2} in the atmosphere into mineral form and longterm storage, provided the surface hosts liquid water.  A more subtle interaction between carbon and oxygen, and the reason for considering them together, is in setting planetary redox.  Carbon's impact on climate is very much dependent on how it is speciated in the atmosphere and in a planet's mantle, both of which depend on how oxidising those respective planetary reservoirs are.  In the mantle, carbon will be present dissolved in nominally carbon-free minerals, and as a separate carbon-bearing phase, carbonate in oxidising mantles and graphite/diamond in reducing mantles.  Changing the mineral form of carbon has a dramatic impact on its release to the environment during magmatism, with reduced forms retaining carbon in the interior more effectively than oxidised forms \citep[e.g.,][]{ortenzi2020_scirep}.  

Carbon is not just passively responding to planetary redox, as an abundant multi-valent element it itself sets how oxidising planetary reservoirs are.  Therefore, the C/O ratio of planetary building blocks is a key variable in deciding the eventual mineralogy, geodynamics, climates, and habitabitabiltiy of planets \citep[e.g.,][]{kuchner2005_arxiv}.  The emerging planets can look qualitative different if a significant component of building blocks condense in disk regions where $\mathrm{C/O}>0.8$, in which case SiC, graphite, and TiC become dominant phases \citep{Bond2010b}. While stellar abundances indicate such ratios are unlikely in systems as a whole (Sect.~\ref{sec:CHNOPS_COSMOS}), disks may develop regions with high C/O ratios as the result of pebble drift (Sect.~\ref{sec:pebble_chnops}). The dominant form of C in carbon-rich planets can vary from graphite and diamond to carbonate rocks when more oxygen is available. The high viscosity and thermal conductivity of diamond in particular (compared to silicates) results in very slow interior convection  \citep{Unterborn2014}. Whilst high C/O planets would have reducing interiors, and more reducing volcanic gasses and surface environments than Earth at any point in its known geological history \citep{catling2020_sciadv}, in some respects they may provide more favourable environments for the origin of life.  In oxidising environments prebiotic chemistry has to compete with oxygen, which will readily react with the reduced compounds life is built out of \citep{Sasselov2020}.  High C/O planets (planets with carbon inventories) will be an exciting target for future habitability studies and observation.

\textit{Hydrogen: }Too much water can pose myriad problems for habitability.  High surface pressure from thick water oceans may reduce a planet's ability to have magmatism, replenish the surface, and drive stabilising feedback cycles. For example, \citet{Kite2009} showed that, for an Earth-mass planet with a stagnant-lid tectonics mode, only 0.4 wt\% water (about 30 Earth oceans) is enough to increase the pressure at the water-rock boundary to prevent decompression melting. This mass fraction decreases with increasing planet mass (to 0.2\% for a $2.5M_\oplus$ planet). Such a lack of melting can shut down deep carbon or water cycles, limiting the planet's ability to regulate its climate. 

Even if a planet is able to regulate its climate with large surface water inventories, and in a sense preserve its habitability, that is not to say it retains the potential for abiogenesis.  Whilst water is a key solvent in prebiotic chemistry, it is critical at stages in reaction pathways to exclude water from the system: for example, phosphorylation reactions require dry-down steps \citep[e.g.,][]{patel2015common}, and it is unclear how these would be achieved in the context of a deep global ocean.  More fundamentally, the emergence of life represents (and likely required) the creation of closed environments in which chemistry could occur isolated from the dilute environment.  The closed basins of sub-aerial land provide an ideal niche for prebiotic chemistry to have occurred on Earth, without the problem of infinite dilution that doing chemistry in a global ocean would entail.

\textit{Nitrogen: }Nitrogen is highly insoluble in magmas, so the likely consequence of increased inventories of N for planets is more massive \ce{N2}-dominated atmospheres.  As \citet{wordsworth2013_science} described, increased nitrogen partial pressure can produce greenhouse warming, and, as with thick water oceans, high pressures of nitrogen would suppress volcanic degassing.  Prebiotic chemistry is reliant on a surficial inventory of nitrogen, which can be incorporated into organic molecules, such as amino acids and nucleotides.  Not only is N a structural component in the key biomolecules of life, but prebiotic chemistry experiments have also identified the N-bearing gas HCN as a key feedstock molecule to initiate prebiotic synthesis: from the earliest spark experiments of Miller-Urey \citep{miller1957_biochim} to the more recent photoredox cyanosulfidic chemistry of \cite{patel2015common}, HCN, and therefore nitrogen is central to building more complex molecules on the way to life.

\textit{Sulfur \& phosporus: }It is less clear what the consequences of large inventories of sulfur and phosphorus would be for chemical habitability.  Phosphorus is a limiting nutrient for life, and perhaps more so for prebiotic chemistry, where it is typically desired in high concentrations.  Simply increasing the abundance of phosphorus would be advantageous in both cases.  Sulfur exists in multiple forms in planets, as gas species in the atmosphere, in dissolved form in oceans, in mineral form in the crust, and as sulfide melts at higher temperature in the mantle.  There are therefore a number of potential reservoirs for sulfur in a planet, where sulfur could be sequestered without impacting habitability. The study of Venus-like exoplanets may provide a key insight into the behaviour of sulfur in planets, and whether sulfur-dominated atmospheres are a common occurrence \citep[e.g.,][]{jordan2021photochemistry}.

\section{SUMMARY}
In conclusion, planet formation is a story of accretion, of elements and compounds with sometimes very different origins coming together and ending up in the same planetary body. But it is also a story of pathways diverging, with tremendous diversity both between planetary systems, and between different worlds orbiting the same star. We have attempted here to provide an overview of the behavior of CHNOPS en route to and during planet formation, highlighting in places the active role CHNOPS-bearing molecules play in these processes. Loss processes and dynamics (i.e., spatial mixing within the protoplanetary disk or in the young planetary system) play a role at virtually every size scale, with pebble drift, planet migration, and dynamical instabilities in particular featuring prominently in modern planet formation theory. Key open questions relevant for understanding a planet's chemical habitability concern e.g., the detailed chemical composition of inner protoplanetary disk midplanes, the timing and location(s) of planetesimal formation, and the thermal evolution of planetary building blocks. During the later stages of planet formation, the continued accretion of pebbles and planetesimals, as shaped by the system's architecture, will have to be considered in concert with the planet's ongoing geophysical evolution. Given the observed variation in pre-stellar, protoplanetary, and mature exo-planetary systems, it seems likely then that small rocky worlds ending up near the habitable zone can form with a range of CHNOPS budgets that exceeds the variation observed in the inner solar system.

\bigskip
\textbf{Acknowledgments.} 
S.~N.~R. is grateful to the CNRS's PNP programming for funding, and to the {\em Virtual Planetary Laboratory} lead team of NASA's Astrobiology Institute.  E.~A.~B. acknowledges support from NASA’s Emerging Worlds Program, grant 80NSSC20K0333, and Exoplanets Research Program, grant 80NSSC20K0259, along with Grant \#1907653 from NSF AAG. M.~K.~M. acknowledges support from the NWO Veni Talent scheme, under the CORVETTE project.
\bigskip


\bibliographystyle{pp7}
\bibliography{main.bib}

\end{document}